\documentclass[fleqn,usenatbib]{mnras}

\usepackage{newtxtext,newtxmath}
\usepackage[T1]{fontenc}

\DeclareRobustCommand{\VAN}[3]{#2}
\let\VANthebibliography\thebibliography
\def\thebibliography{\DeclareRobustCommand{\VAN}[3]{##3}\VANthebibliography}

\usepackage{graphicx}	
\usepackage{amsmath}	
\usepackage[version=4]{mhchem}
\usepackage{gensymb}
\usepackage{chemfig}

\let\oldAA\AA
\renewcommand{\AA}{\text{\normalfont\oldAA}}

\title[Clouds on HAT-P-1b]{\centering{Modelling dynamically driven global cloud formation microphysics in the HAT-P-1b atmosphere.}}

\author[Elspeth K.H. Lee]{
Elspeth K.H. Lee$^{1}$ \\
$^{1}$ Center for Space and Habitability, University of Bern, Gesellschaftsstrasse 6, CH-3012 Bern, Switzerland
}

\date{Accepted XXX. Received YYY; in original form ZZZ}

\pubyear{2023}

\begin{document}
\label{firstpage}
\pagerange{\pageref{firstpage}--\pageref{lastpage}}
\maketitle

\begin{abstract}
Insight into the formation and global distribution of cloud particles in exoplanet atmospheres continues to be a key problem to tackle going into the JWST era.
Understanding microphysical cloud processes and atmospheric feedback mechanisms in 3D has proven to be a challenging prospect for exoplaneteers.
In an effort to address the large computational burden of coupling these models in 3D simulations, we develop an open source, lightweight and efficient microphysical cloud model for exoplanet atmospheres.
`Mini-cloud' is a microphysical based cloud model for exoplanet condensate clouds that can be coupled to contemporary general circulation models (GCMs) and other time dependent simulations.
We couple mini-cloud to the Exo-FMS GCM and use a prime JWST target, the hot Jupiter HAT-P-1b, as a test case for the cloud formation module.
After 1000+ of days of integration with mini-cloud, our results show a complex 3D cloud structure with cloud properties relating closely the dynamical and temperature properties of the atmosphere.
Current transit and emission spectra data are best fit with a reduced cloud particle number density compared to the nominal simulation, with our simulated JWST NIRISS SOSS spectra showing promising prospects for characterising the atmosphere in detail.
Overall, our study is another small step in first principles 3D exoplanet cloud formation microphysical modelling.
We suggest that additional physics not included in the present model, such as coagulation, are required to reduce the number density of particles to appropriately observed levels.
\end{abstract}

\begin{keywords}
planets and satellites: atmospheres -- planets and satellites: individual: HAT-P-1b -- planets and satellites: gaseous planets
\end{keywords}



\section{Introduction}

Clouds are a ubiquitous part of atmospheres inside and outside the Solar System, dictating a significant or sometimes dominant part of their observational properties.
In the JWST Early Release Science (ERS) transiting exoplanet program \citep{Batalha2017,Bean2018} transit spectra of the hot Saturn planet WASP-39b was observed across a variety of instrument modes \citep{Ahrer2023, Alderson2023, Feinstein2023,Rustamkulov2023}, finding that models with a cloud component were best able to fit the data across the wide wavelength range of all observing modes.
\citet{Sing2016} produced a holistic look at the spectra of ten hot Jupiters with HST and Spitzer data, finding a continuum of cloudy spectral features dependent on the equilibrium temperature of the planets.
Clouds are a prime candidate to explain the nightside brightness temperature trends for hot Jupiter exoplanets \citep{Beatty2019,Keating2019}, where the nightside brightness temperatures remain clustered around 1000 K for planets below equilibrium temperatures of around 2000 K. 

Modelling observable trends in cloudy exoplanets has been attempted by a few studies.
In \citet{Gao2020}, a combined cloud and haze model was used to fit the amplitude trend found in the 1.4$\mu$m \ce{H2O} feature in transmission spectra that is affected by aerosol particle opacity.
\citet{Gao2021} model the cloud profiles of the nightside regions of hot Jupiters, concluding that the Spitzer brightness temperatures can be explained through changes in the cloud top pressure as a function of the planetary equilibrium temperature.

In concert with recent JWST data, high resolution spectroscopy from ground based instrumentation such as ESPRESSO and HARPS are now able to probe the condensation fronts of refractory species in ultra hot Jupiters (UHJs) \citep[e.g.][]{Ehrenreich2020, Kesseli2021} and more generally discover chemical gradients between the east and west terminator regions in UHJ atmospheres \citep[e.g.][]{Sluijs2023}.
This new JWST and high-resolution era warrants a three dimensional (3D) understanding of the atmospheric distribution of cloud particles, due to the fidelity and spacial information shown by these observatories providing detailed insight into the chemical composition and cloud properties of exoplanets hemisphere to hemisphere.

A popular option to model clouds across an exoplanet atmosphere is to post-process output of a GCM model and apply a 1D cloud formation code to extracted cloud free vertical T-p structures.
This has been performed successfully in many studies \citep[e.g.][]{Lee2015,Powell2018,Helling2021,Robbins-Blanch2022}, giving detailed insight into global distributions of cloud structures and the expected cloud compositions across a wide range of exoplanet system parameters.

The microphysical cloud formation bin method model CARMA has been used in this context, where in \citet{Powell2018} they processed outputs from the SPARC/MITgcm \citep{Showman2009, Parmentier2016} across a wide equilibrium temperature range.
This model was also used in \citet{Gao2020} and \citet{Gao2021} to explain trends in cloud affected transmission spectra features seen as a function of equilibrium temperature.

Most commonly, phase equilibrium models, where the cloud particles are assumed to be at equilibrium with the surrounding gas vapour, have been coupled to 3D GCM models of exoplanet atmospheres.
For example, \citet{Charnay2015} use a simple phase equilibrium conversion scheme to model KCl and ZnS clouds in GJ 1214b.
The one dimensional EddySed \citep{Ackerman2001} model has been coupled into the Met-Office UM GCM to model clouds on HD 209458b \citep{Lines2019, Christie2021} and GJ 1214b \citep{Christie2022}.
\citet{Roman2017,Roman2019,Roman2021} apply a temperature dependent cloud opacity scheme with prescribed cloud distributions and sizes which is fed-back into the GCM radiative-transfer (RT) module.
Similarly, \citet{Parmentier2021} use a temperature dependent cloud opacity scheme to mimic the radiative effects of cloud formation.
\citet{Komacek2022} use a vapour and condensate tracer scheme with a relaxation timescale method coupled to the gas replenishment and settling rate to convert between both quantities.
A common property of the above phase equilibrium models is the assumption of a particular size distribution of particles, arranging the condensed mass into corresponding parameterised size-distribution parameters (e.g. assuming a mean particle size and variance), generally a log-normal distribution is the distribution of choice due to its useful mathematical properties.

Only a couple of studies to date have attempted to couple microphysical based cloud formation models to large scale hydrodynamic simulations.
These have used the DIHRT model \citep{Lee2016}, with \citet{Lee2016} coupling to the radiative-hydrodynamic atmospheric model of \citet{Dobb-Dixon2013}, and \citet{Lines2018} coupling DIHRT to the UK Met Office UM GCM \citep{Mayne2014}.
Microphysical models typically attempt to simulate the complete lifecycle of the cloud size-distribution time-dependently in some manner, nucleating seed particles from the gas phase and considering the condensation and evaporation rates of individual species with time. 
This adds considerable complexity and additional physics to consider compared to phase equilibrium assumptions.
In addition, typically size-distributions are organically computed in microphysical models rather than imposed.
However, such models have proven to be extremely computationally intensive, with the above studies only able to run for 100 simulated days coupled to the cloud formation model, with substantial high performance computing resources required to be utilised.

To address the issue of computational feasibility of microphysical models we develop a new, open source and more efficient microphysical cloud formation model, `mini-cloud'.
We perform 3D GCM simulations of the hot Jupiter HAT-P-1b as an example testbed of our new cloud formation module.

HAT-P-1b is a hot Jupiter exoplanet discovered by \citet{Bakos2007}.
It is a prime candidate for atmospheric characterisation due to its inflated radius (1.32 R$_{\rm Jup}$) and reduced bulk density due to its mass of 0.53 M$_{\rm Jup}$ \citep{Sing2016}.
\citet{Wakeford2013} observed a transit of HAT-P-1b using WFC3 onboard HST, finding a strong \ce{H2O} signature in the near-IR, indicative of a relatively cloud free atmosphere.
\citet{Nikolov2014} used the HST STIS instrument to measure the optical wavelength transmission spectrum, finding a strong Na feature but a lack of K absorption.
They find that the wings of the Na feature are cut off, suggesting a cloud opacity component as a possibility to fit and connect the STIS and WFC3 data consistently.
This HST data was collated in the \citet{Sing2016} summary of ten hot Jupiters as well as the addition of Spitzer 3.6 $\mu$m and 4.5 $\mu$m transit depths.
\citet{Barstow2017} perform retrieval modelling of the HAT-P-1b data from \citet{Sing2016} finding grey or Rayleigh scattering clouds between 0.1 and 0.01 bar are consistent with the data.
\citet{Pinhas2019} suggest a super-solar \ce{H2O} abundance with a median cloud top pressure of 0.1 bar when using their retrieval framework on the same data.
HAT-P-1b is scheduled to be observed with the JWST NIRISS SOSS instrument for transit and eclipse during Cycle 1 \citep{Lafreniere2017} which will add further detailed characterisation information on the atmosphere.

In Section \ref{sec:mc} we present the mini-cloud model as well as how it differs from the full DIHRT model.
In Section \ref{sec:model} we detail the expected cloud formation species on HAT-P-1b and details on the GCM setup with mini-cloud.
Section \ref{sec:res} presents the results of our coupled 3D GCM modelling of HAT-P-1b and the produced cloud structures.
In Section \ref{sec:pp} we post-process our GCM results to produce synthetic transmission and secondary eclipse emission observables.
Section \ref{sec:dis} contains the discussion of our results and Section \ref{sec:con} contains the summary and conclusion.

\section{mini-cloud}
\label{sec:mc}

Mini-cloud is an open source\footnote{\url{https://github.com/ELeeAstro/mini_cloud}} microphysical cloud formation model that has its origins in the methodologies of Helling and collaborators \citep[DRIFT:][]{Woitke2003,Helling2008}, which in turn is based on the theory and development from modelling AGB star wind dust formation \citep[e.g.][]{Gail2013}.

A time-dependent version of the underlying theory was utilised in \citet{Lee2016}, DIHRT, to be coupled to hydrodynamic simulations of exoplanet atmospheres.
However, a major limitation of this model was the inability to simulate beyond 100 simulated Earth days due to the large computational effort required to couple the full microphysical model \citep{Lines2018}.
With this background, we develop mini-cloud as an offshoot of DIHRT, designed to be much more efficient at coupling to 3D GCMs and allowing longer integration timescale simulations to be performed.
However, this comes at the price of reduced complexity compared to the full DIHRT model (Section \ref{sec:dihrt_diff}).
Mini-cloud can therefore be seen as an intermediate complex model for simulating cloud microphysics in large scale projects where computational efficiency is paramount.

Mini-cloud uses the `method of moments' to evolve the integrated cloud particle size-distribution. 
The ith moment of the particle size-distribution, K$_{i}$ [cm$^{i}$ cm$^{-3}$], is defined as \citep{Gail2013}
\begin{equation}
    K_{i} = \int^{\infty}_{a_{0}}a^{i}f(a)da,
\end{equation}
where a$_{0}$ [cm] is the seed particle size and f(a) [cm$^{-3}$ cm$^{-1}$] the particle size distribution as a function of particle size a [cm].
The moments therefore represent integrated quantities of the size distribution, for example, the total cloud particle number density, N$_{0}$ [cm$^{-3}$], is equal to the zeroth moment
\begin{equation}
N_{0} = K_{0},
\end{equation}
the number density weighted mean grain size, $\left<a\right>$ [cm], is
\begin{equation}
\label{eq:amean}
\left<a\right> = \frac{K_{1}}{K_{0}}.
\end{equation}
Another useful quantity is the area weighted mean grain size, also known as the effective radius, $a_{\rm eff}$ [cm], given by the ratio
\begin{equation}
\label{eq:aeff}
 a_{\rm eff} = \frac{K_{3}}{K_{2}}.
\end{equation}
The mean particle area, $\left<A\right>$ [cm$^{2}$], and volume $\left<V\right>$ [cm$^{3}$] are then
\begin{equation}
  \left<A\right> = 4\pi\frac{K_{2}}{K_{0}},
\end{equation}
and
\begin{equation}
\label{eq:vmean}
  \left<V\right> = \frac{4\pi}{3}\frac{K_{3}}{K_{0}},
\end{equation}
respectively.

To integrate the moments in time, we use the equation set described in \citet{Gail2013} where the time derivative of each moment (from i=0-3) is given by
\begin{align}
\frac{d K_{0}}{dt} &= J_{*} + J_{\rm evap},\\
\frac{d K_{1}}{dt} &= a_{0}(J_{*} + J_{\rm evap})  + \frac{da}{dt}K_{0}, \\
\frac{d K_{2}}{dt} &= a_{0}^{2}(J_{*} + J_{\rm evap}) + 2\frac{da}{dt}K_{1}, \\
\frac{d K_{3}}{dt} &= a_{0}^{3}(J_{*} + J_{\rm evap}) + 3\frac{da}{dt}K_{2}.
\end{align}
Following \citet{Helling2008}, this equation set is supplemented with a K$_{3}$ moment evolved for each cloud species s 
\begin{equation}
\frac{d K_{3}^{s}}{dt} = a_{0}^{3}(J_{*}^{s} + J_{\rm evap}^{s}) + 3\frac{da^{s}}{dt}K_{2},
\end{equation}
which tracks the individual contribution of each cloud species to the total volume of the particle distribution.
By definition, the total K$_{3}$ value is given by the sum of each individual species
\begin{equation}
    K_{3} = \sum_{s}K^{s}_{3}.
\end{equation}
Due to this, mini-cloud is a mixed species `dirty' grain model the same way as the DRIFT model \citep[e.g.][]{Helling2008} with multiple condensed species contributing simultaneously to the grain bulk properties and composition. 

For the nucleation rate, J$_{*}$ [cm$^{-3}$ s$^{-1}$], we use the modified classical nucleation theory as presented in several sources \citep[e.g.][]{Gail2013, Helling2013,Lee2018}. 
A single specific species is chosen as the main seed particle carrier (\ce{TiO2}, C, SiO, KCl or NaCl in mini-cloud).
For this specific HAT-P-1b case, we chose \ce{TiO2} with the updated temperature dependent surface tension expression from \citet{Sindel2022}.
Throughout mini-cloud, a seed particle size of 1 nm is assumed.

Unlike in previous studies, we include explicitly the rate of evaporation of seed particles, J$_{\rm evap}$ [cm$^{-3}$ s$^{-1}$], into the differential equation set.
This is justified as in a typical particle size distribution a population of seed particles can be present, which would evaporate for a given atmospheric thermochemical condition.
For example, a significant population of seed particles can be seen in the CARMA simulations of \citet{Powell2018}, suggesting that accounting for their evaporation time dependently is an important consideration.
The seed particle evaporation rate is estimated to be
\begin{equation}
    J^{s}_{\rm evap} = \frac{K_{0}}{a_{0}}\frac{da^{s}}{dt},
\end{equation}
which only applies when da$^{s}$/dt $<$ 0.

The cloud particle growth or evaporation rate, da/dt [cm s$^{-1}$], for a species $s$, is given by \citep{Gail2013}
\begin{equation}
   \frac{da^{s}}{dt} = V_{0}^{s}\alpha^{s} n_{i}^{s} \sqrt{\frac{k_{b}T}{2\pi m_{i}}}S,
\end{equation}
where $\alpha$ the reaction efficiency, m$_{i}$ [g] the mass of the reactant and V$_{0}^{s}$ [cm$^{3}$] the unit volume of the condensed species monomer. 
S is the stability coefficient given by
\begin{equation}
\label{eq:stab}
    S = \left(1 - \frac{1}{S_{\rm r}}\right),
\end{equation}
where S$_{\rm r}$ $\approx$ p$_{\rm par}^{s}$/p$_{\rm vap}^{s}$, the ratio of the partial pressure to the species vapour pressure is the reaction supersaturation ratio of the cloud species \citep{Helling2006}.
Following \citet{Woitke2020}, when S is negative it is multiplied by the volume mixing ratio of the cloud species to the total bulk, K$_{3}^{s}$/K$_{3}$.

The consumption rate of the number density of condensable gas, i, n$_{i}$ [cm$^{-3}$], due to dust species s is given by \citep{Gail2013}
\begin{equation}
\frac{dn_{i}}{dt} = - N_{l}^{s}(J_{*}^{s} + J_{\rm evap}^{s}) - \frac{4\pi K_{2}}{V_{0}^{s}}\frac{da^{s}}{dt},
\end{equation}
where N$_{l}^{s}$ is the number of monomers that make up one seed particle size.
To integrate the equation set in time we use the stiff ODE solver dvode which is part of the odepack\footnote{\url{https://computing.llnl.gov/projects/odepack}} package.

\subsection{The Kelvin effect}
\label{sec:KE}

An important consideration when modelling the condensation of species onto a surface is the Kelvin effect in which a condensing droplet more efficiency grows on a flat surface compared to a curved surface.
This is represented by the dimensionless factor, K$_{\rm f}$,
\begin{equation}
\label{eq:Kf}
    K_{\rm f} = \exp\left({\frac{2\sigma^{s}_{\infty}V_{0}^{s}}{ak_{b}T}}\right),
\end{equation}
where $\sigma^{s}_{\infty}$ [erg cm$^{-2}$] is the bulk surface tension of the condensate.
This alters the stability criteria from Eq. \ref{eq:stab}, which then becomes
\begin{equation}
    S = \left(1 - \frac{K_{\rm f}}{S_{\rm r}}\right).
\end{equation}
The Kelvin effect therefore increases the supersaturation required by a condensation species before condensation will occur, depending on the bulk properties of the material through the surface tension and the radius of the surface the species is condensing on.
The K$_{\rm f}$ factor strongly depends on the size of the surface, becoming negligible ($\sim$ 1) for large grains.
The Kelvin effect is therefore most important when considering condensation and evaporation on smaller grain sizes.
In mini-cloud we assume the grain size to be the effective particle size a$_{\rm eff}$ in Eq. \ref{eq:Kf}.

\subsection{Cloud particle opacity}

As in DIHRT, we use Mie theory to calculate the opacity and scattering properties of the cloud particles. 
The optical constants of the mixed grain are calculated using Landau-Lifshitz-Looyenga effective medium theory \citep{Looyenga1965}, which combines each individual species optical constants weighted by their volume mixing ratio contribution to the grain bulk.
Optical constants for each species are taken from the \citet{Kitzmann2018} collection. 

We use the LX-MIE routine from \citet{Kitzmann2018} to perform the Mie theory calculations.
To increase computational efficiency and avoid expensive Mie calculations, for small size parameters (x $<$ 0.01) we use the Rayleigh scattering limit approximation \citep{Bohren1983}.
For large size parameters (x $>$ 100) we use the Anomalous Diffraction Approximation (ADT) \citep[e.g.][]{Hoffman2016}.
Therefore, the Mie theory calculations are only applied for intermediate size parameters (0.01 $<$ x $<$ 100), which increases the computational efficiency of the opacity calculation scheme considerably compared to using Mie theory across the whole size parameter range.
However, this comes at the cost of some accuracy, particularly for large size parameter regimes where the real refractive index is large (n $\gtrsim$ 1.5), which is common for the mineral materials used in this study.
The calculated cloud opacity is then mixed with the gas phase opacities for use inside the RT scheme.

\subsection{Cloud particle settling}

\begin{table}
\centering
\caption{Eq. \ref{eq:rosner} parameters for different species useful for hydrogen dominated atmospheres.}
\begin{tabular}{c c c c}  \hline
Species & d [$\AA$] & $\epsilon_{\rm LJ}$ [erg] & m [g] \\ \hline
\ce{H2} & 2.827 & 59.7 $k_{\rm b}$ & 2.016 amu \\
He & 2.511 & 10.22 $k_{\rm b}$ & 4.003 amu \\
H & 2.5 & 30.0 $k_{\rm b}$ & 1.008 amu \\ \hline
\end{tabular}
\label{tab:rosner}
\end{table}%

To describe the settling of cloud particles, we follow the Stokes flow prescription presented in \citet{Parmentier2013}, where the terminal fall speed, $v_{\rm f}$ [cm s$^{-1}$], is given as
\begin{equation}
\label{eq:vf}
v_{\rm f} = -\frac{2\beta a^{2}g(\rho_{\rm d} - \rho_{\rm gas})}{9\eta} ,
\end{equation}
where $a$ [cm] is the cloud particle radius, g [cm s$^{-2}$] the local vertical gravitational acceleration (here taken as a positive quantity), $\rho_{\rm d}$ [g cm$^{-3}$] the bulk density of the cloud particle and $\rho_{\rm gas}$ [g cm$^{-3}$] the density of the local gas phase.
$\beta$ is the Cunningham slip factor
\begin{equation}
\beta = 1 + K_{N}\left\{1.256 + 0.4\exp(-1.1/K_{N})\right\} ,
\end{equation}
with $K_{N}$ the Knudsen number
\begin{equation}
K_{N} = \frac{\lambda}{a},
\end{equation}
where $\lambda$ [cm] is the atmospheric mean free path (Eq. \ref{eq:path_length}).

The dynamical viscosity, $\eta$ [g cm$^{-1}$ s$^{-1}$], of the background gas is given by \citet{Rosner2012}
\begin{equation}
\label{eq:rosner}
\eta = \frac{5}{16}\frac{\sqrt{\pi m k_{b}T}}{\pi d^{2}}\frac{(k_{b}T/\epsilon_{\rm LJ})^{0.16}}{1.22},
\end{equation}
where $d$ [cm] is the molecular diameter, $m$ [g] the molecular mass and $\epsilon_{\rm LJ}$ [K] the depth of of the Lennard-Jones potential.
For the background gases of interest in this study, \ce{H2} and He, we take the parameters from \citet{Rosner2012}, replicated in Table \ref{tab:rosner}.

For an ideal gas, the mean free path, $\lambda$ [cm], in each layer is related to the dynamical viscosity from
\begin{equation}
\label{eq:path_length}
\lambda =  \frac{\eta}{p} \sqrt{\frac{\pi k_{b} T}{2 \bar{\mu} \ \textrm{amu}}},
\end{equation}
where $\bar{\mu}$ [g mol$^{-1}$] is the mean molecular weight of the background gas.

We assume that the particle size used in Eq. \ref{eq:vf} is the effective particle size (Eq. \ref{eq:aeff}).
In the model Eq. \ref{eq:vf} is calculated each dynamical timestep and converted to pressure velocity units (Pa s$^{-1}$) assuming hydrostasy.
We then use a MacCormack method vertical advection routine with minmod flux limiter to calculate the vertical flux of the cloud particle moment solutions.
In testing, it was found the vertical advection was a prime source of numerical instability and produce nonphysical cloud particle properties. 
It was important to rigorously impose all boundary conditions and set the local vertical velocity to zero should the number density of cloud particles be less than 10$^{-10}$ cm$^{-3}$.
Future studies will investigate different methodologies and more sophisticated vertical advection schemes to achieve greater accuracy, however in practice our current scheme produces satisfactory and stable results.
Horizontal and vertical advection of cloud particle tracers due to atmospheric flows is handled directly by the GCM dynamical core.

\subsubsection{Viscosity of gas mixtures}

We include the ability to specify a static background mixture with individual species mixing ratios.
To capture the effect of gas mixtures on the dynamical viscosity, we apply the classical viscosity mixing law \citep{Wilke1950}
\begin{equation}
\label{eq:Wilke}
\eta_{\rm mix} = \sum_{i=1}^{N}\frac{y_{i}\eta_{i}}{\sum_{j=1}^{N}y_{i}\psi_{ij}},
\end{equation}
where $\eta_{\rm mix}$ [g cm$^{-1}$ s$^{-1}$] is the dynamical viscosity of the gas mixture and $y_{i}$ the molar ratio of component $i$.
The function $\psi_{ij}$ is given by
\begin{equation}
\psi_{ij} = \frac{\left[1 + \sqrt{\frac{\eta_{i}}{\eta_{j}}}\sqrt[4]{\frac{m_{j}}{m_{i}}}\right]^{2}}{\frac{4}{\sqrt{2}}\sqrt{1 + \frac{m_{i}}{m_{j}}}},
\end{equation}
where $m$ [g] is the molecular mass of the gas species.
The path length of the mixture can then be calculated from Eq. \ref{eq:path_length}.
For simplicity we assume a constant neutral gas Solar \ce{H2} and He ratio of 0.85 and 0.15 respectively.

\subsubsection{Importance of considering gas mixtures}

\begin{figure} 
   \centering
   \includegraphics[width=0.48\textwidth]{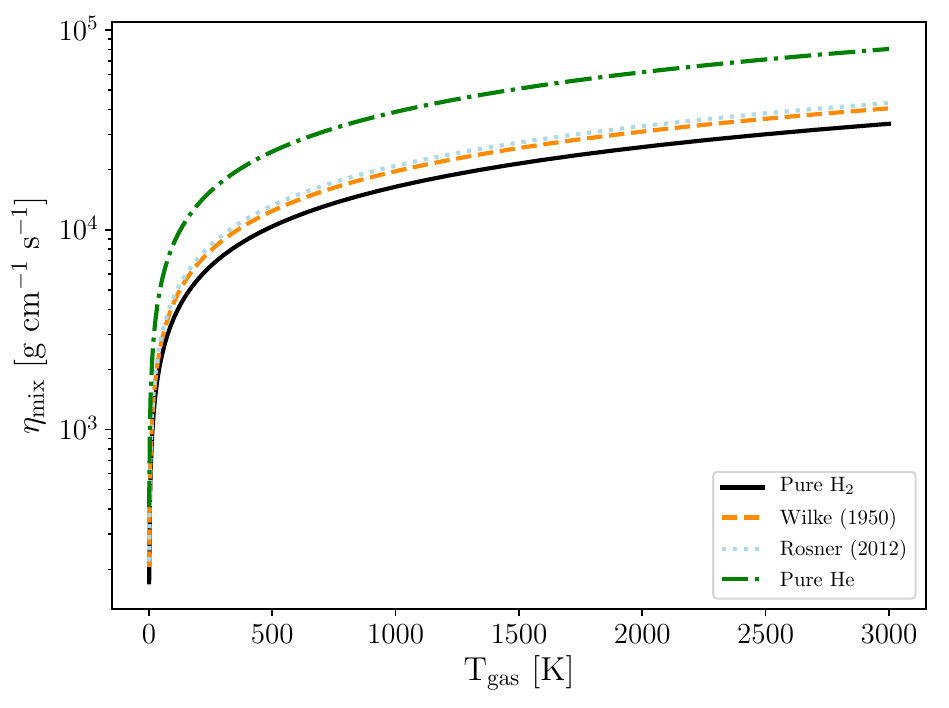}
   \caption{Comparison of the dynamical viscosity mixing rules (Eqs. \ref{eq:Wilke} and \ref{eq:Rosner}) for a 85\%-15\% \ce{H2}-He background gas mixture, compared to the pure components.}
   \label{fig:eta_mix}
\end{figure}

We briefly examine the importance of considering \ce{H2} and He gas mixtures compared to their pure components.
Additionally, we include the approximate `square root' mixing law \citep{Rosner2012}
\begin{equation}
\label{eq:Rosner}
\eta_{mix} = \frac{\sum_{i} \sqrt{m_{i}} y_{i} \eta_{i}}{\sum_{i} \sqrt{m_{i}} y_{i}}.
\end{equation}

Figure \ref{fig:eta_mix} presents the dynamical viscosity of pure \ce{H2} and He and a 0.85 - 0.15 mix, typical of a neutral gas solar metallicity mixing ratio, using Eqs \ref{eq:Wilke} and \ref{eq:Rosner}.
The difference between the pure \ce{H2} and mixtures range between 10-15\% dependent on the temperature.
With the viscosity of the mixture being larger than the pure \ce{H2} case, it can be expected that the settling velocity from Eq. \ref{eq:vf} will be lower, barring any possible cancelling out effects, for the mixture than that to pure \ce{H2}.
In addition, from Figure \ref{fig:eta_mix} the \citet{Rosner2012} square root expression approximates the full \citet{Wilke1950} equation well, suggesting the square root expression is a simple way to take into account background mixtures when required.

\subsection{Differences to DIHRT}
\label{sec:dihrt_diff}

In this section, we detail some of the key differences between mini-cloud and the DIHRT model.
Mini-cloud is a simplified version of DIHRT, with several approximations and assumptions compared to the full model.

In DIHRT several gas to solid phase reactions (sometimes in the 10s \citet{Helling2008}) for each species are considered to contribute to the growth or evaporation of the grain surface.
In mini-cloud this is replace with a single pseudo-reaction where the fictitious condensable gas is directly condensed into the solid phase. 
For example, 
\begin{equation}
    \ce{Mg2SiO4} \rightarrow \ce{Mg2SiO4}[s],
\end{equation}
where $[s]$ denotes the solid phase.
The initial abundance of the fictitious gas may be set to the limiting elemental abundance at some metallicity ratio, e.g. Mg at $\approx$ 3.548$\cdot$10$^{-5}$ for the \citet{Asplund2021} solar atomic ratios, dividing by the stoichiometric factor of that element in the pseudo-gas (i.e. 2 for Mg).
The stoichiometric factor of the pseudo-reactions are therefore always equal to one, simplifying the equation set further.
This is similar to the concepts used in phase equilibrium modelling \citep[e.g.][]{Morley2012}, where a representative atomic abundance is used as a proxy for condensing molecules.
As a consequence, a large difference between the DIHRT and mini-cloud model is that mini-cloud does not perform any chemical equilibrium calculations during integration, rather directly condensing the representative element.
This simplification is the most important time-saving measure of mini-cloud compared to DIHRT.
However, this means mini-cloud may over-represent the rate of growth of cloud particles, as the rate is not tempered by considering the changing rates of each individual surface chemical reaction due to changes in gas phase compositions, or by any temperature and pressure dependence on the chemical equilibrium abundance of gas species that take part in said surface reactions.

A particular numerical instability occurs in both DIHRT and mini-cloud when strong evaporation of cloud particles occurs across a rapid timescale, this can lead to overshooting the saturation limits and oscillating behaviour, resulting in overestimating the gas phase abundances from the evaporation process.
In our experience, this behaviour is only triggered in deep regions when large cloud particles rapidly fall past their thermal stability conditions.
To counter this DIHRT utilised an instant evaporation technique \citep{Lee2016} to ease the timescale of evaporation on the ODE solver.
However, this method adds large computational expense as the threshold of instant evaporation has be checked in small timesteps during integration to be successful.
In mini-cloud, we instead assume that the maximum abundance of evaporated material cannot exceed the deep atmosphere initial conditions. 
This is probably a reasonable assumption as typically the instability regions occur deep in the atmosphere, where convective motions would rapidly homogenise the atmospheric tracers.

\section{Modelling the atmosphere of HAT-P-1b}
\label{sec:model}

In this section, we detail the approach to modelling the HAT-P-1b atmosphere coupled to the mini-cloud scheme.

\subsection{Expected clouds on HAT-P-1b}

\begin{figure}
    \centering
    \includegraphics[width=0.49\textwidth]{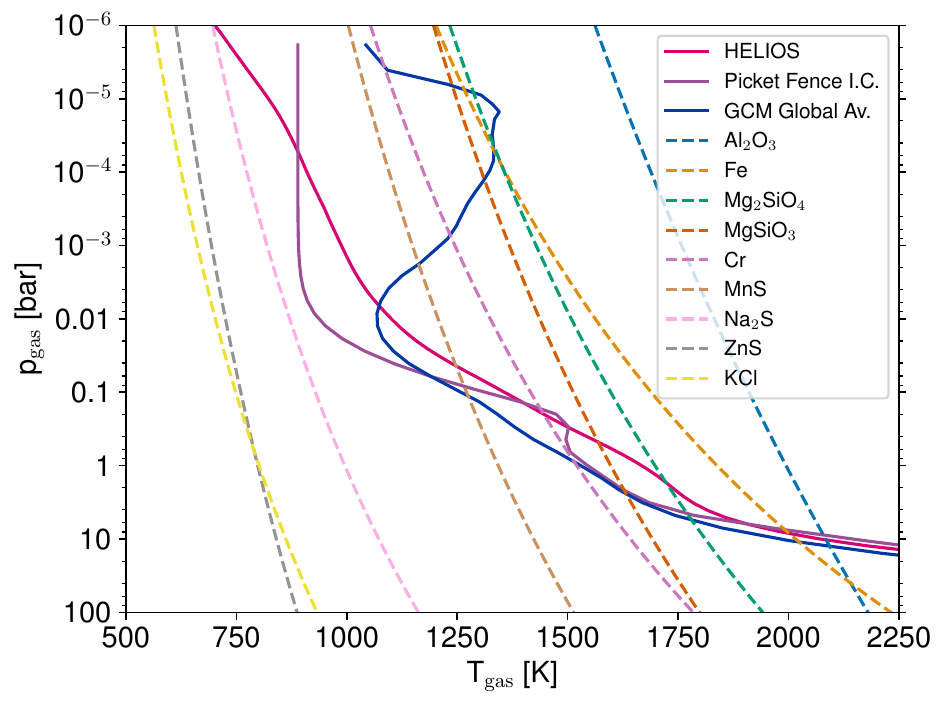}
    \caption{HELIOS RCE, picket-fence GCM initial conditions (IC) and globally average GCM T-p profile of the HAT-P-1b atmosphere (black dashed line) compared to the saturation limit curves of various cloud forming species.}
    \label{fig:HELIOS}
\end{figure}

\begin{figure*}
    \centering
    \includegraphics[width=0.49\textwidth]{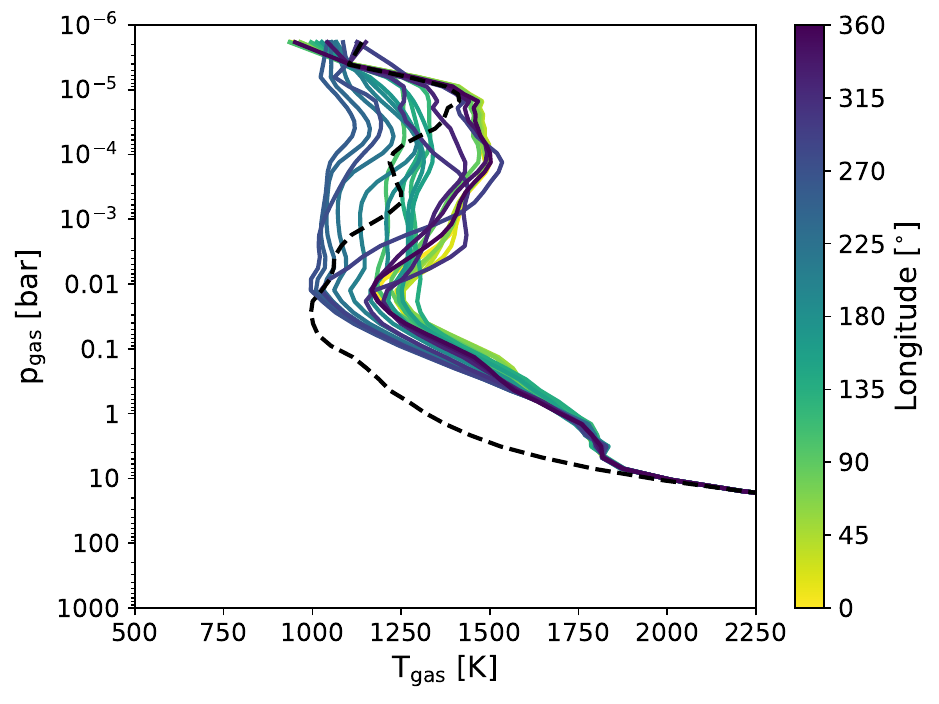}
    \includegraphics[width=0.49\textwidth]{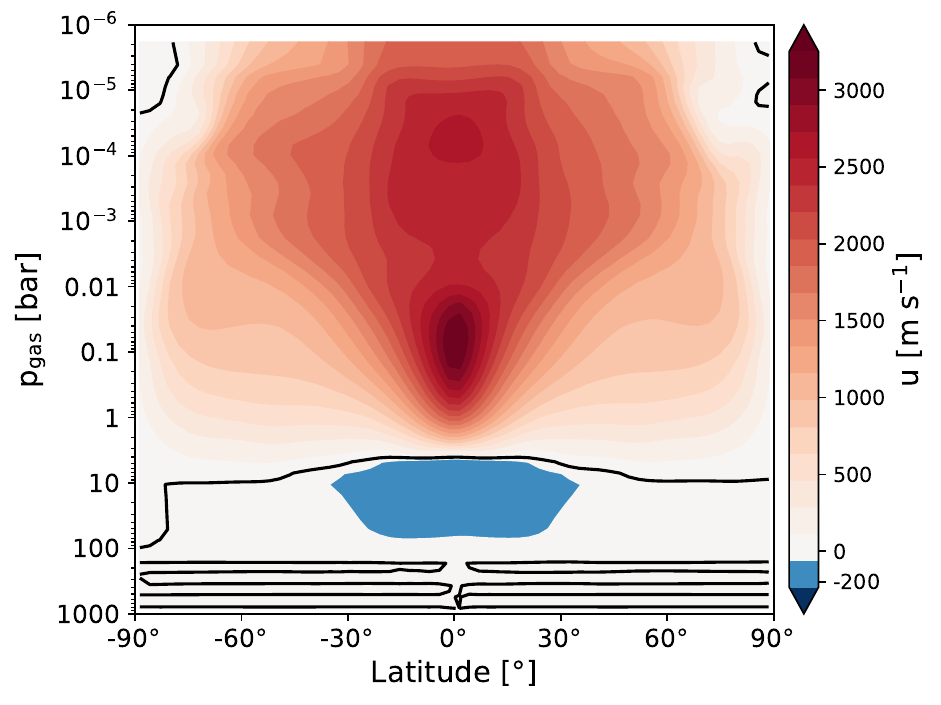}
    \includegraphics[width=0.49\textwidth]{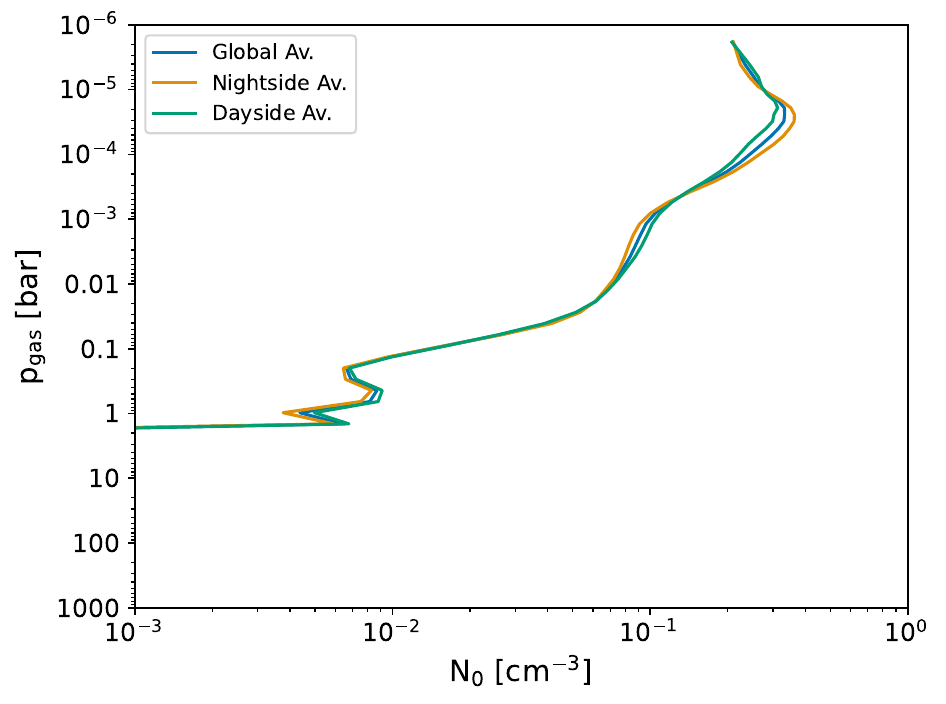}
    \includegraphics[width=0.49\textwidth]{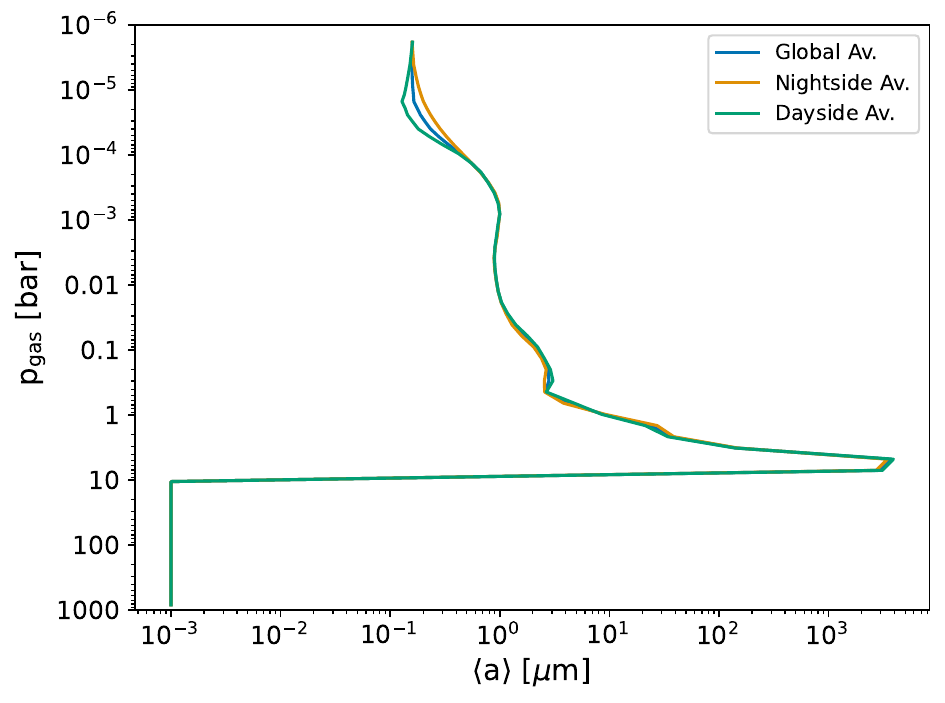}
    \caption{Results from the Exo-FMS GCM model with mini-cloud. 
    Top left: vertical temperature-pressure profiles at the equator region (coloured lines) and polar region (dashed line).
    Top right: zonal mean of the zonal velocity component of the wind.
    Bottom left: global, nightside and dayside averaged cloud particle number density.
    Bottom right: global, nightside and dayside averaged mean cloud particle size.}
    \label{fig:GCM}
\end{figure*}

As a first estimate of the expected cloud composition found in HAT-P-1b we perform 1D radiative-convective equilibrium (RCE) calculations using the HELIOS code \citep{Malik2017,Malik2019} with system parameters taken from \citet{Nikolov2014}.
We assume a full globe redistribution factor of 0.25.
We then overplot the thermochemical stability curves of various condensed species that have been published in the literature \citep{Visscher2010,Morley2012,Wakeford2017}.

Figure \ref{fig:HELIOS} shows the result of this initial 1D test.
We can see that the main condensate species to consider are \ce{TiO2}, \ce{Al2O3}, \ce{Fe}, \ce{Mg2SiO4}, \ce{MgSiO3}, \ce{Cr} and \ce{MnS}.
Due to their high surface tensions, Cr and MnS are unlikely to condense in significant mass for this case \citep{Gao2020}.
Therefore, for simplicity we simulate \ce{TiO2}, \ce{Al2O3}, \ce{Fe} and \ce{Mg2SiO4}, opting for one silicate species as representative of the silicate clouds component to simplify the calculations. 
We assume \ce{TiO2} as the primary seed particle nucleating species.

\subsection{GCM modelling}

\begin{table}
\centering
\caption{Adopted GCM simulation parameters for HAT-P-1b. We use a cubed-sphere resolution of C32 ($\approx$ 128 $\times$ 64 in longitude $\times$ latitude).}
\begin{tabular}{c c c l}  \hline \hline
 Symbol & HAT-P-1b  & Unit & Description \\ \hline
 T$_{\rm int}$ & 477 & K & Internal temperature \\
 T$_{\rm eq}$ & 1325 & K & Eq. temperature \\
 P$_{\rm 0}$ & 1000 &  bar & Reference surface pressure \\
 c$_{\rm P}$ & 12822  &  J K$^{-1}$ kg$^{-1}$ & Specific heat capacity \\
 R & 3577 &  J K$^{-1}$ kg$^{-1}$  & Ideal gas constant \\
 $\kappa$ & 0.279 & -  & Adiabatic coefficient \\
 g$_{\rm p}$ & 7.46  & m s$^{-2}$ & Acceleration from gravity \\
 R$_{\rm p}$ & 9.43 $\times$ 10$^{7}$  & m & Radius of planet \\
 $\Omega_{\rm p}$ & 1.63 $\times$ 10$^{-5}$ & rad s$^{-1}$ & Rotation rate of planet \\
 $\Delta$ t$_{\rm hyd}$ & 30  & s & Hydrodynamic time-step \\
 $\Delta$ t$_{\rm rad}$  & 300 & s & Radiative time-step \\
 $\Delta$ t$_{\rm cld}$  & 300 & s & mini-cloud time-step \\
 N$_{\rm v}$ & 54  & - & Vertical resolution \\
 d$_{\rm 4}$ & 0.16  & - & $\mathcal{O}$(4) div. dampening coefficient \\
\hline
\end{tabular}
\label{tab:GCM_parameters}
\end{table}

We use the Exo-FMS GCM model in gas giant configuration \citep{Lee2021} which has been used to model a variety of exoplanet regimes, from sub-Neptunes \citep{Innes2022} to ultra hot Jupiters \citep{Lee2022b}.
Our adopted GCM parameters are shown in Table \ref{tab:GCM_parameters}, derived from the \citet{Nikolov2014} measurements.
We use the stellar parameters of HAT-P-1 from \citet{Nikolov2014} (T$_{\rm eff}$ = 5980 K, log g = 4.359, [M/H] = 0.13) and use a PHOENIX stellar atmosphere model \citep{Husser2013} to calculate the flux incident onto the planetary atmosphere in each wavelength band.

For our radiative-transfer solution, the longwave radiation is calculated using the Toon source function technique \citep{Toon1989}, a 1D two-stream multiple scattering method commonly used across many exoplanet GCM models \citep[e.g.][]{Showman2009, Roman2017}.
For the shortwave radiation we use the adding method \citep{Mendonca2015}, which takes into account the scattering of incident starlight.
In this way we can consistently take into account the scattering feedback of the clouds, either from direct scattering of incident starlight or scattering of thermal radiation inside the atmosphere.

We use a correlated-k opacity scheme with 11 bands as in \citet{Kataria2013} with pre-mixed k-tables the same as in \citet{Lee2022b}, but without including strong optical and UV wavelength absorbing species such as TiO, VO, Fe and SiO.
Rayleigh scattering by \ce{H2} and He as well as collisional induced absorption by \ce{H2}-\ce{H2} and \ce{H2}-He pairs is included.

With this scheme we do not include feedback of cloud particle formation on the gas phase composition through changes in the refractory and oxygen element abundance by the condensation and evaporation processes. 
For example, the condensation of \ce{Mg2SiO4} would reduce the oxygen atoms available in the gas phase by $\approx$ 10\% compared to solar ratios, which will affect the the abundance of oxygen bearing species such as \ce{H2O}.
Therefore, our current set-up is not fully self-consistent in this regard and is a key aim for future iterations of this model.
The changes to C/O ratio and chemical composition of this effect is explored in detail in \citet{Helling2021,Helling2023} across a wide range of planetary parameters.

\subsubsection{Initial conditions and spin-up}

We use an internal temperature of 477 K, derived using the expression in \citet{Thorngren2019}.
The \citet{Parmentier2014,Parmentier2015} picket-fence scheme is used to generate the initial T-p profile conditions of the GCM model (Figure \ref{fig:HELIOS}).
As in \citet{Lee2022b}, we perform an extended 2000 day spin-up period using the picket-fence RT scheme before switching to the correlated-k scheme for 1000 additional days.
We then run the correlated-k scheme coupled to mini-cloud for 2000 days but neglect the effects of radiative-feedback of cloud opacity to allow some evolution of the cloud scheme.
Finally, radiative-feedback with the cloud particle opacity included is run for 500 days to complete the simulation, with the first 100 days consisting of a ramping up period (see next paragraph).
An additional 50 days is then performed, the average results of which is taken as the final product.

To retain numerical stability during the phase where cloud feedback is first included we follow a similar scheme to \citet{Lines2018}, linearly increasing the cloud opacity over the course of 100 days of simulation.
As in \citet{Lee2016}, we also apply a vertical boxcar smoothing of the cloud opacity properties to reduce the effect of spikes on opacity (for example, regions at the top or bottom of the cloud layer) that can produce too large heating gradients for the simulation to handle.

Unfortunately, during runtime it was discovered that too strong thermal instabilities occur in the model where the cloud opacities change too rapidly in the vertical direction. 
Typically this occurs from a combination of large cloud opacity gradients with large single scattering albedo and asymmetry parameters, notorious conditions where two-stream RT schemes traditionally struggle with.
To combat this, we limit the single scattering albedo and asymmetry factor to a maximum of 0.75 and divide the extinction opacity by 50.
We also limit the cloud opacity to a minimum of 10$^{-5}$ bar, we find including cloud opacity at the uppermost layers led to huge instability with the RT scheme, probably due to the extremely small radiative-timescales at these pressures.
Though these limitations are significant, we find this scheme to be adequate in accounting for the large scale radiative-feedback effects of the cloud particle formation.
Improvements in this area will be investigated in future iterations of the model.

For the mini-cloud initial conditions, we assume a cloud-free atmosphere. 
We assume a deep reservoir of condensable gas by setting the initial abundance of the gas tracers at their values where the pressure of the atmosphere is greater than 10 bar, slightly below the saturation temperature limit of \ce{Al2O3}. 
Every timestep, this deep region is removed of cloud particles and replenished to the initial gas abundances, emulating break up of cloud particles through convective motions accompanied by a strong replenishment of condensable materials.
This pressure is also within the radiative-convective boundary, where we can expect tracers to be well mixed due to convective motions.
This significantly reduces the time that would be required to mix enough condensable gas from only the lower boundary to the saturation point of \ce{TiO2}, which we estimate would take 1000s of days of simulation.
Using our initial condition set-up with the \citet{Asplund2021} solar atomic abundances, during testing it was found to take around five simulated days before cloud formation was triggered due to upward mixing of the deep reservoir of condensable gas.
This scheme is significantly different to the initial conditions used in \citet{Lee2016} and \citet{Lines2018}, where a Solar abundance across the globe was assumed as initial conditions. 
The current scheme aims to emulate the `bottom up' approach used in cloud formation models such as EddySed \citep{Ackerman2001} and CARMA \citep{Gao2020}, where condensable gas is mixed upwards to the condensation layers from a deep gas reservoir.

\section{Cloud structures of HAT-P-1b}
\label{sec:res}

\begin{figure*}
    \centering
    \includegraphics[width=0.49\textwidth]{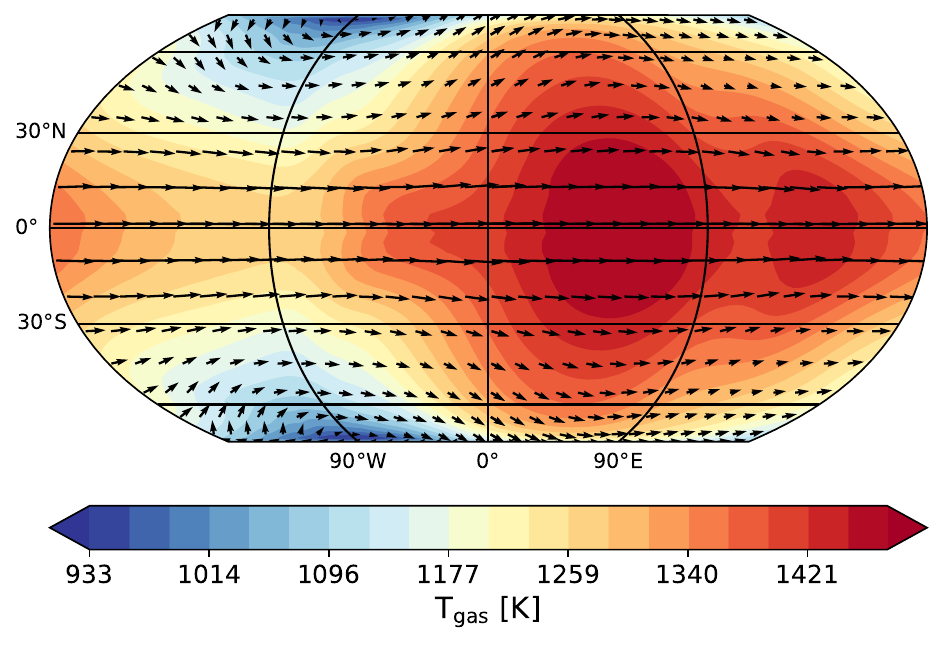}
    \includegraphics[width=0.49\textwidth]{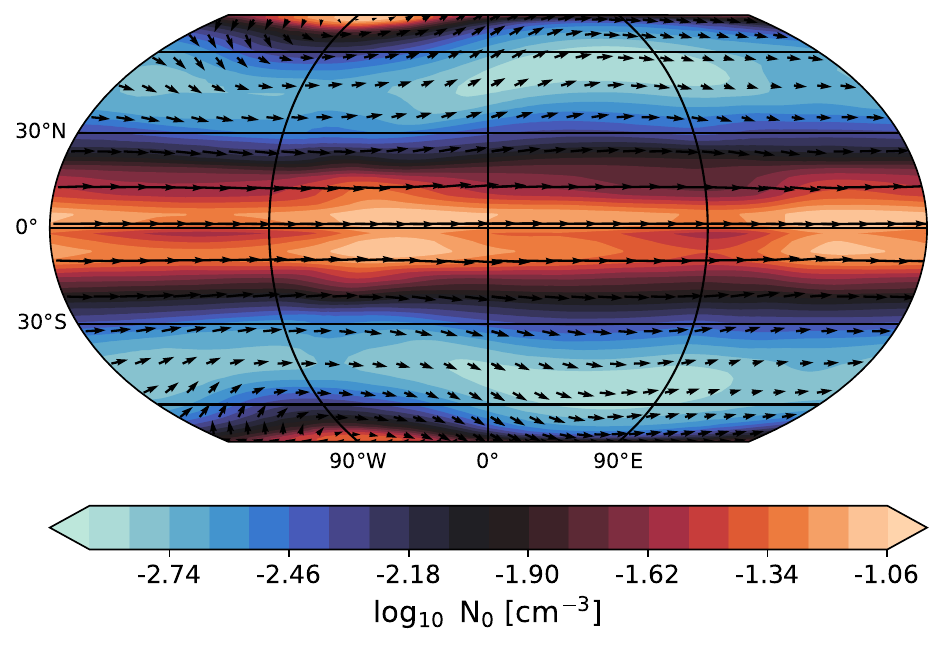}
    \includegraphics[width=0.49\textwidth]{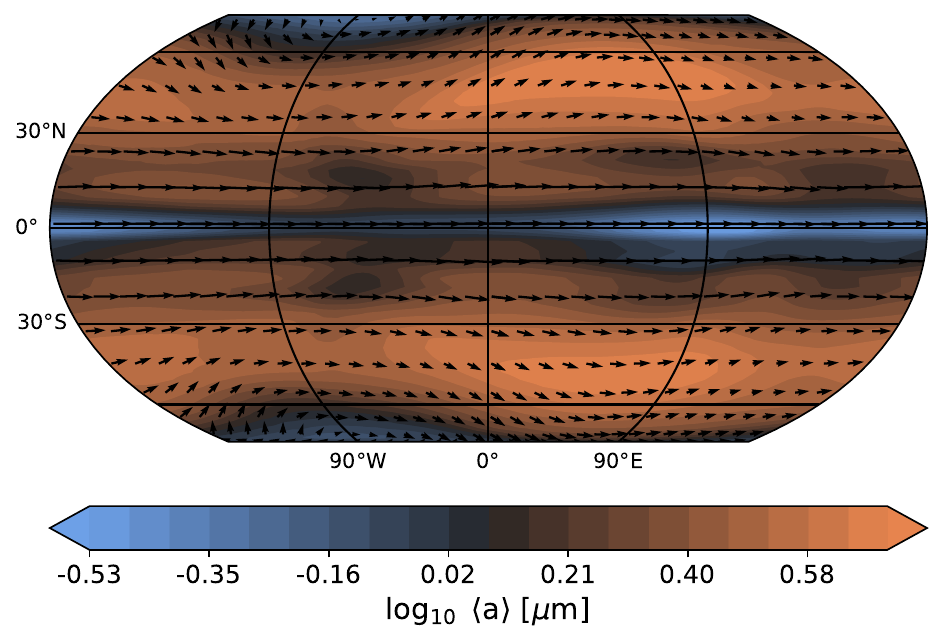}
    \includegraphics[width=0.49\textwidth]{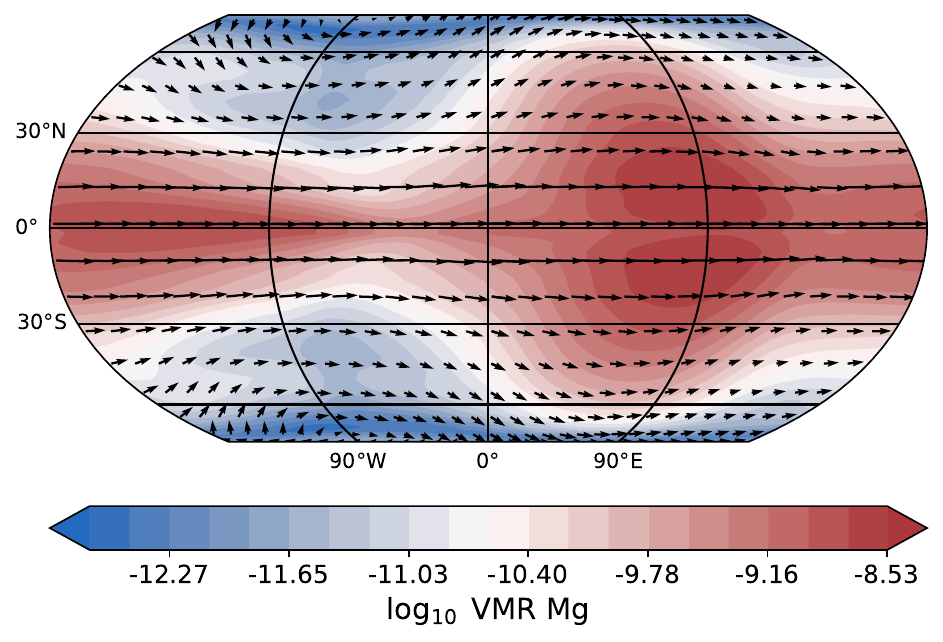}
    \caption{Longitude-latitude maps at the 0.1 bar pressure level of the GCM model. The sub-stellar point is located at 0$\degree$,0$\degree$.  Top left: Temperature of the GCM. Top right: Number density of the cloud particles. Bottom left: Effective particle size.
    Bottom right: volume mixing ratio of Mg condensable gas material.}
    \label{fig:map}
\end{figure*}

\begin{figure*}
    \centering
    \includegraphics[width=0.49\textwidth]{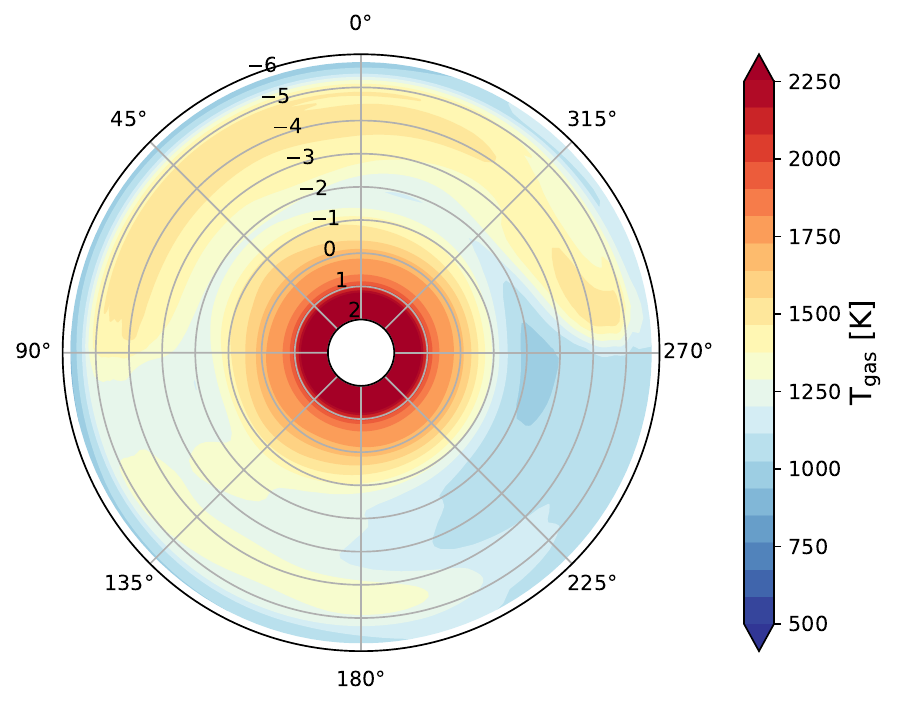}
    \includegraphics[width=0.49\textwidth]{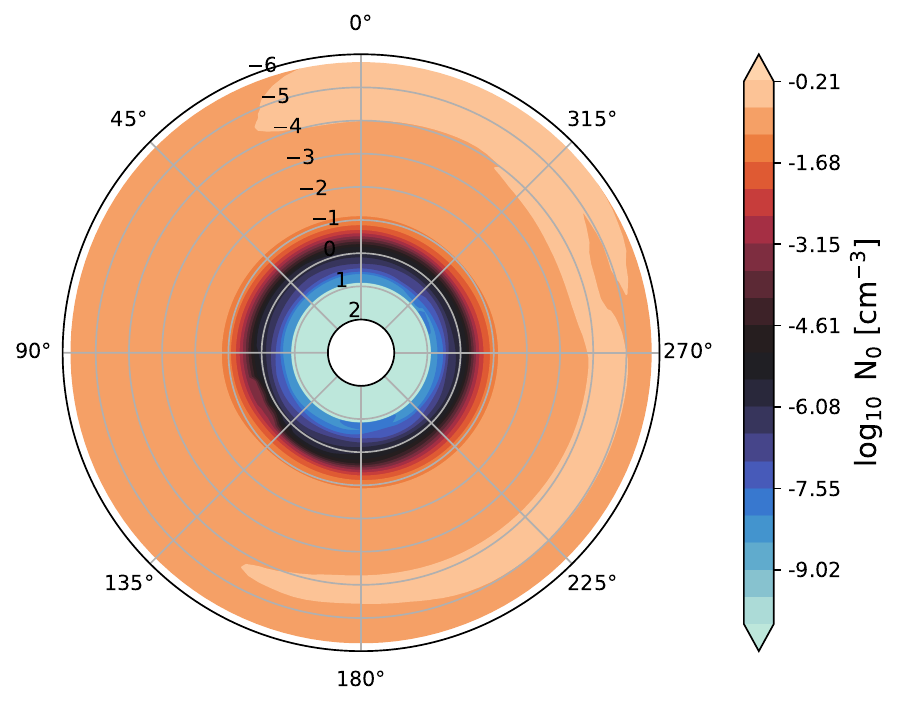}
    \includegraphics[width=0.49\textwidth]{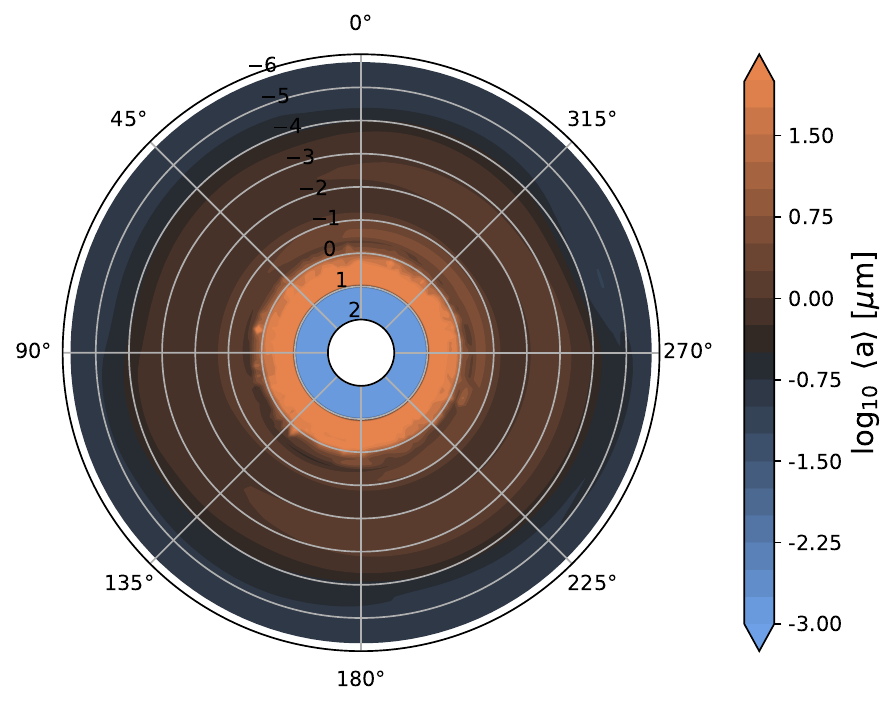}
    \includegraphics[width=0.49\textwidth]{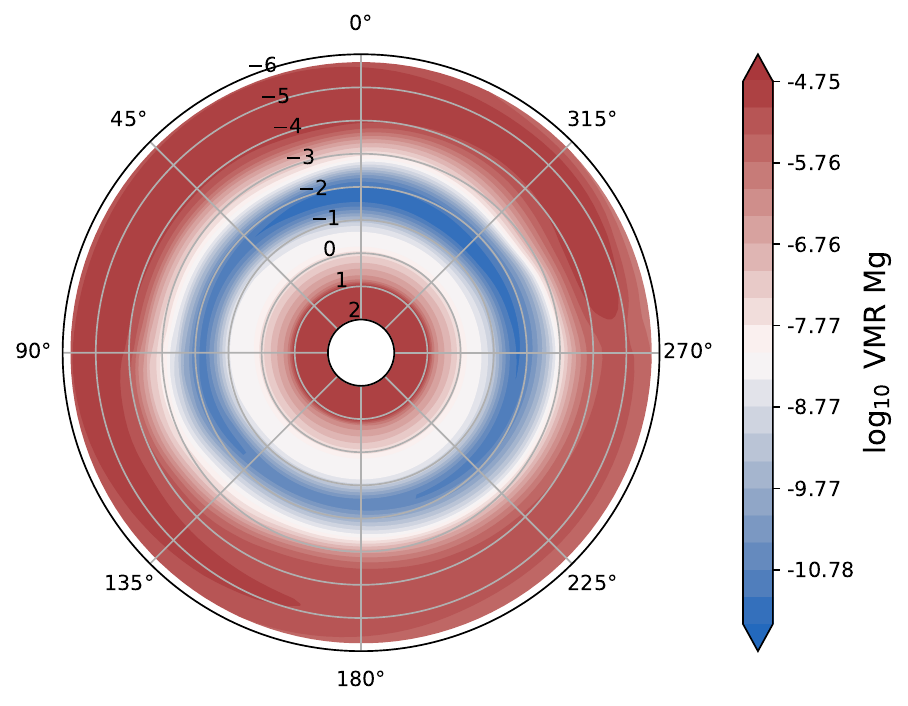}
    \caption{Longitude coordinate temperature and cloud property maps from the GCM averaged across the central jet region $\pm$20$\degree$ latitude from 0$\degree$ latitude. 0$\degree$ shows the sub-stellar region. The radial numbers denote the pressure level in $\log_{10}$ bar. Top left: Temperature of the GCM. Top right: Number density of the cloud particles. Bottom left: Effective particle size.
    Bottom right: volume mixing ratio of Mg condensable gas material.}
    \label{fig:meri}
\end{figure*}

\begin{figure*}
    \centering
    \includegraphics[width=0.49\textwidth]{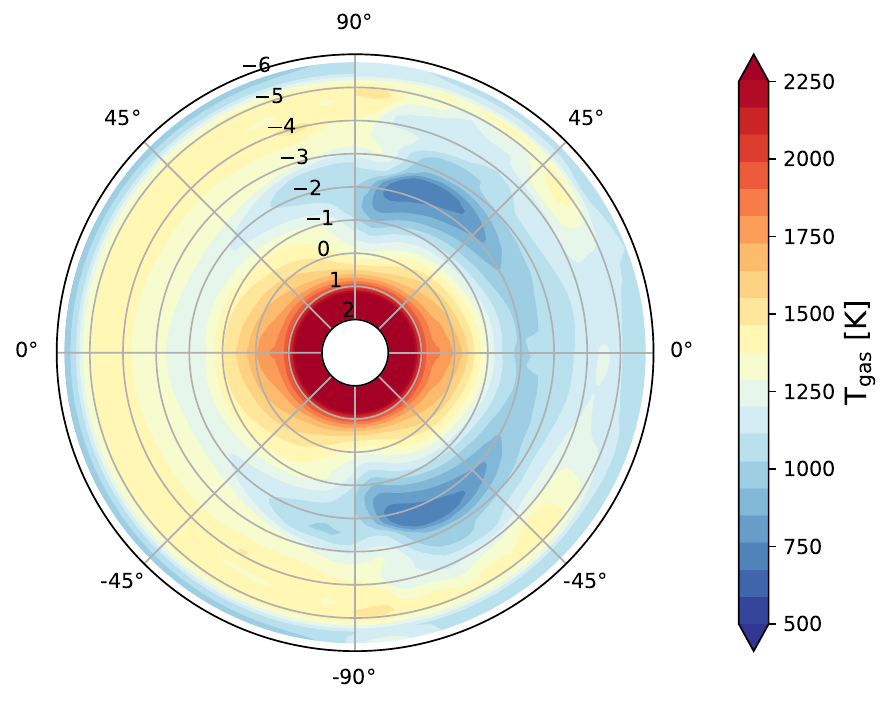}
    \includegraphics[width=0.49\textwidth]{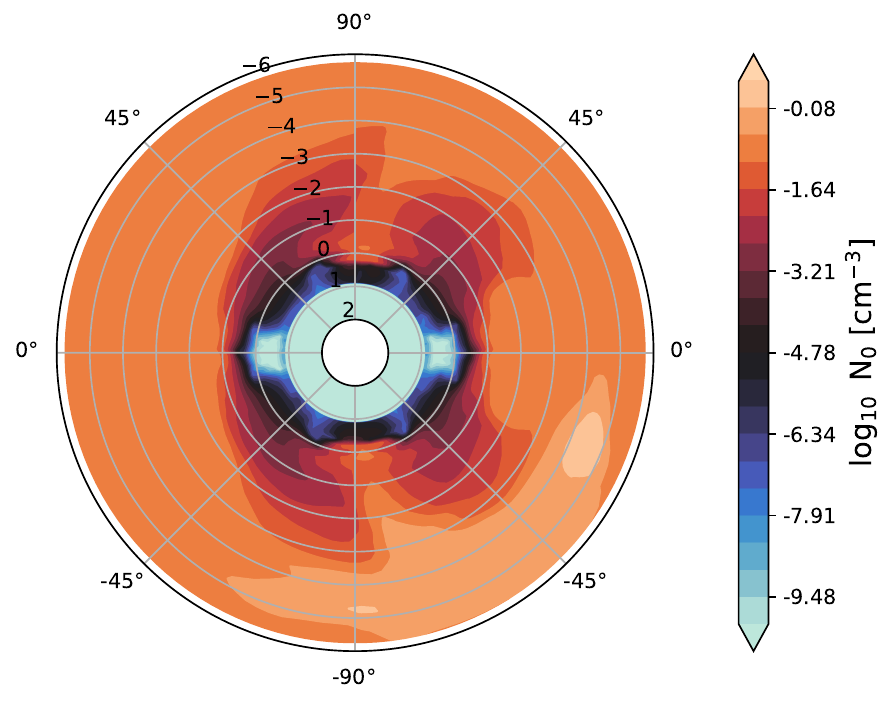}
    \includegraphics[width=0.49\textwidth]{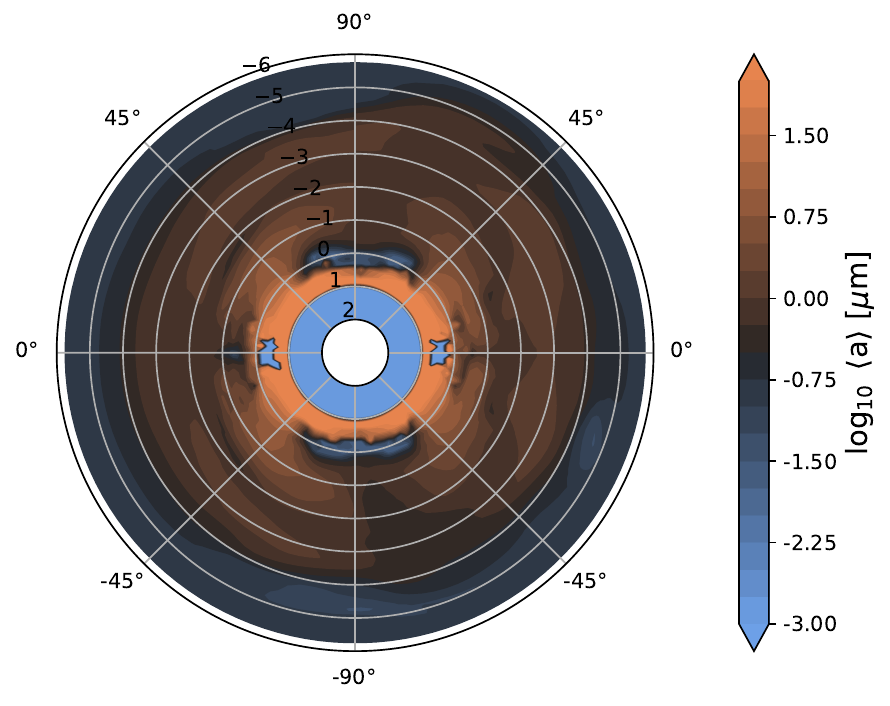}
    \includegraphics[width=0.49\textwidth]{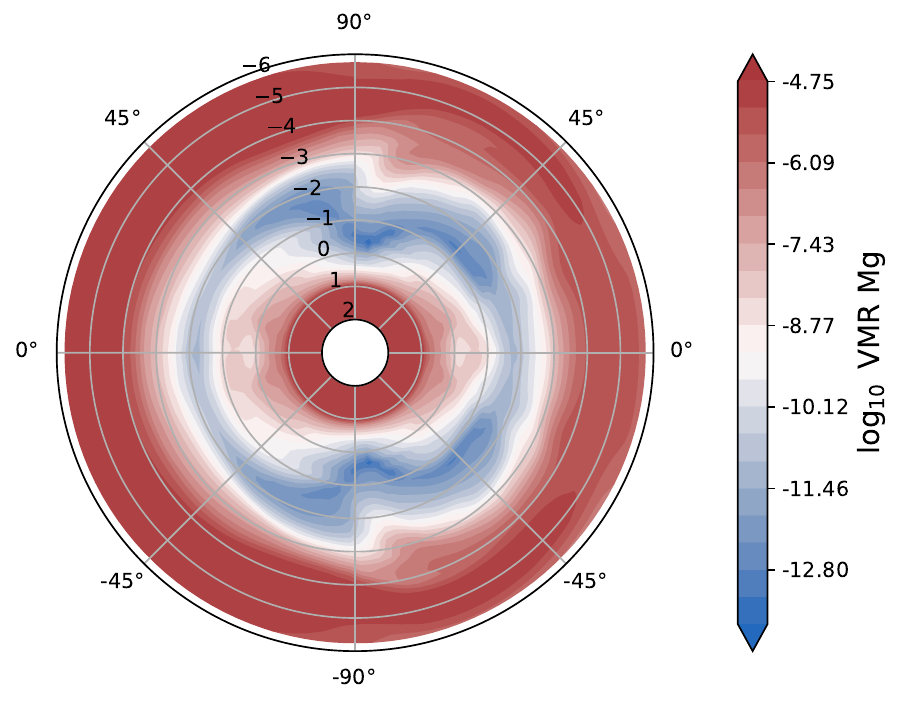}
    \caption{Latitude coordinate temperature and cloud property maps from the GCM averaged across the approximate opening angle $\pm$10$\degree$ longitude from the east (left half of plots) and west (right half of plots) terminators. The radial numbers denote the pressure level in $\log_{10}$ bar. Top left: Temperature of the GCM. Top right: Number density of the cloud particles. Bottom left: Effective particle size.
    Bottom right: volume mixing ratio of Mg condensable gas material.}
    \label{fig:trans}
\end{figure*}

\begin{figure*}
    \centering
    \includegraphics[width=0.49\textwidth]{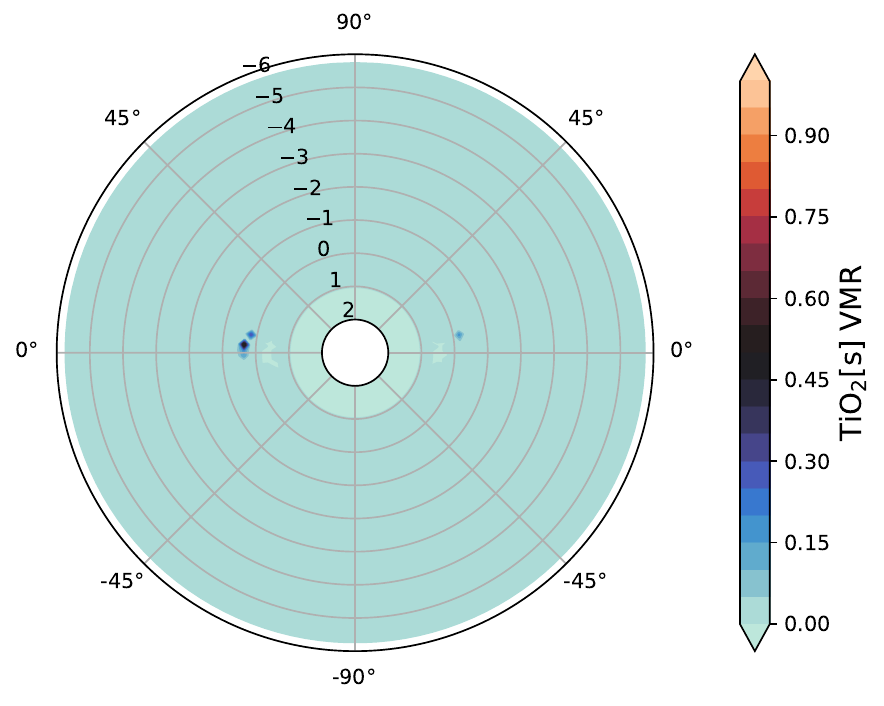}
    \includegraphics[width=0.49\textwidth]{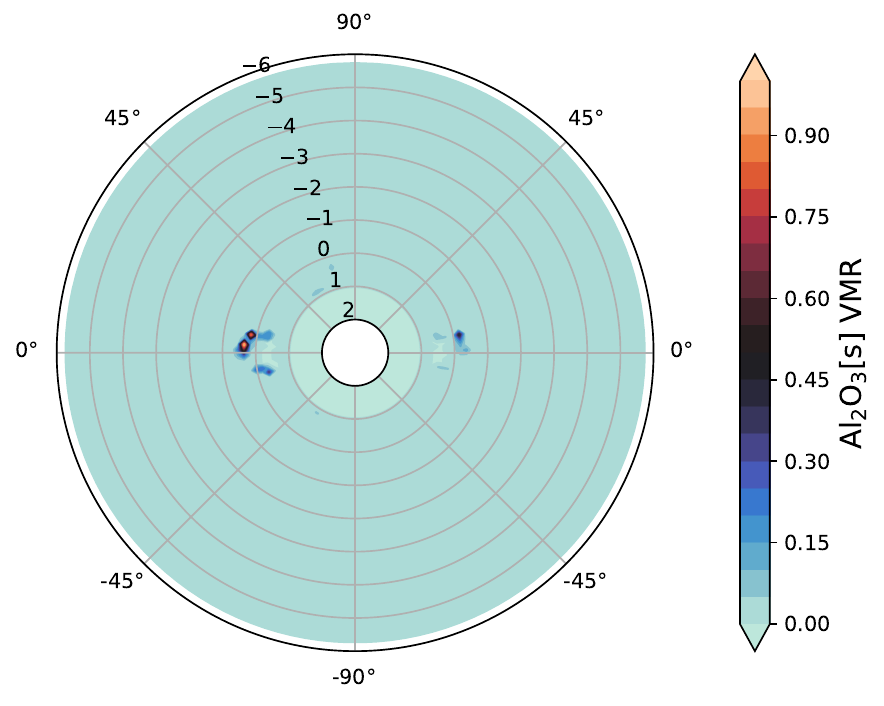}
    \includegraphics[width=0.49\textwidth]{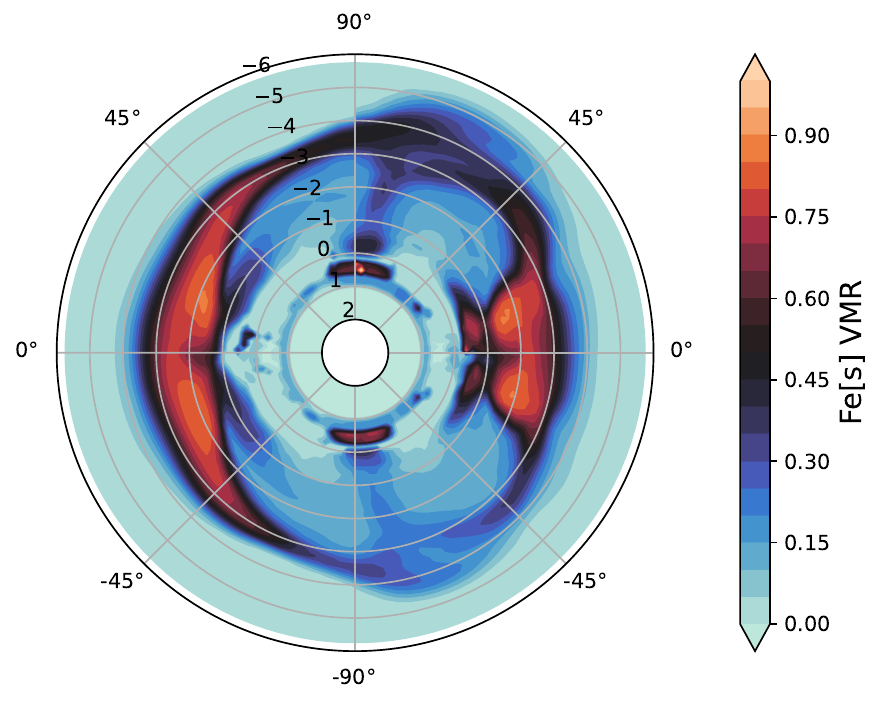}
    \includegraphics[width=0.49\textwidth]{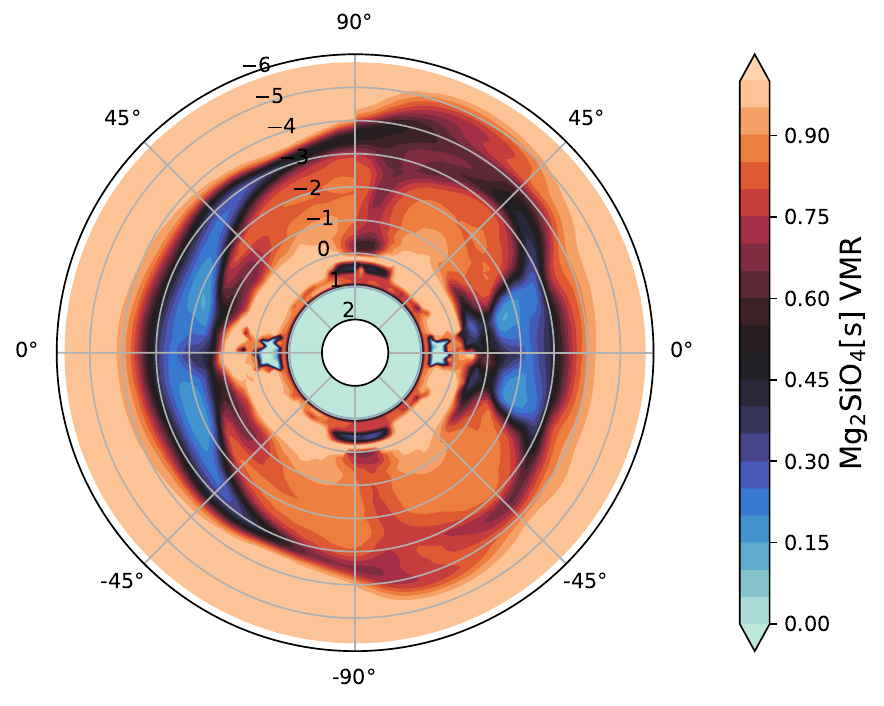}
    \caption{Latitude coordinate maps of the cloud particle bulk volume mixing ratio of each condensed species from the GCM averaged across the approximate opening angle $\pm$10$\degree$ longitude from the east (left half of plots) and west (right half of plots) terminators. The radial numbers denote the pressure level in $\log_{10}$ bar. Top left: \ce{TiO2}[s] volume mixing ratio. Top right: \ce{Al2O3}[s] volume mixing ratio. Bottom left: \ce{Fe}[s] volume mixing ratio. Bottom right: \ce{Mg2SiO4}[s] volume mixing ratio.}
    \label{fig:trans_bmix}
\end{figure*}

In Figure \ref{fig:GCM} we present the vertical temperature-pressure (T-p) structures and zonal mean zonal velocity of the HAT-P-1b GCM model with mini-cloud and Figure \ref{fig:HELIOS} shows the globally averaged T-p profile from the GCM compared to the HELIOS and picket-fence initial conditions.
This shows a highly typical HJ transport pattern for this dynamical regime, with a strong and wide equatorial jet.
Clouds have greatly affected the upper atmospheric temperature structure, inducing an inversion at low pressures and more isothermal-like atmosphere in these regions.
This is most apparent in Figure \ref{fig:HELIOS}, where the globally average GCM T-p profile shows a large deviation in the upper atmospheric profiles compared to the HELIOS and picket-fence models.
We also see influences deeper in the atmosphere, the temperature shows a stark uptick when clouds begin to form in the atmosphere around 1-10 bar and add their opacity to the atmosphere.
The zonal mean zonal velocity shows a typical wind pattern for hot Jupiters seen in many hot Jupiter GCM studies, with a strong central jet region. 
These structures suggest two main equatorial jets, one at around the 0.1 bar level and one at low pressure at around 10$^{-4}$ bar.

In Figure \ref{fig:GCM} we also show global, nightside and dayside averaged cloud number density and average particle size vertical profiles. 
This shows that both the global cloud properties of the atmosphere are generally homogeneous between the nightside and dayside hemisphere, suggesting strong transport and mixing in the zonal direction.

Figure \ref{fig:map} presents latitude-longitude maps of the GCM output at the 0.1 bar pressure level. 
This shows a typical HJ thermal structure, with the hottest points offset east from the sub-stellar point.
The cloud particle properties follow closely the dynamical structure of the atmosphere, the largest number densities of cloud particles in the atmosphere are found at the equatorial regions and colder nightside high latitude Rossby gyres.
These Rossby gyres are key nucleation zones of new cloud particles, as also seen in \citet{Lee2016} and \citet{Lines2018}.
This pressure level contains mostly micron and sub-micron sized particles, split between sub-micron at the equator and micron sizes at latitude.
This result shows that the spacial variations in cloud properties are mostly with latitude than longitude, with quite homogeneous cloud properties at each longitude of the planet.
This suggests that the cloud particle properties strongly follow the global flow and wave patterns.
The depletion of Mg atoms clearly follows the temperature profile of the planet in this case.

Figure \ref{fig:meri} presents the results of the GCM at the equatorial jet region averaged across $\pm$20$\degree$ latitude.
Here, the clear differences between the dayside and nightside temperature structures can be seen. 
The number density is highly uniform in the upper atmosphere, but decreases sharply at pressures greater than 10$^{-1}$ bar where seed particles begin to be thermally unstable.
This is reflected in the particle sizes, which shows mostly micron or sub-micron particles mixed across the globe but larger particles deeper than around 1 bar.
The Mg abundance shows a ring structure from around 1 to 10$^{-3}$ bar, suggesting a Mg stability zone has formed at these pressures.

According to the calculations performed in \citet{Wardenier2022}, the opening angle of HAT-P-1b is around 20$\degree$ (H$_{\rm p}$ $\approx$ 635 km, R$_{\rm p}$ $\approx$ 14.7 R$_{\rm E}$).
Therefore, for the transmission limb contour plots in Figure \ref{fig:trans} we average across $\pm$ 10$\degree$ at the east and west terminators.
These plots show the latitudinal structure of the GCM results, with the temperature plot showing the difference between the east and west terminators. 
The cloud structures show large variations with latitude, here we see that the high latitudes contain more cloud particle number density and subsequently slightly smaller particle sizes.
We also see that the latitude of these reduced number density regions is different between the east and west terminator.
Again, the Mg abundance shows a stability zone similar to the equatorial plots, suggesting a spherical shell of \ce{Mg2SiO4} stability is present across the globe of the planet.

\subsection{Cloud composition}

In Figure \ref{fig:trans_bmix} we repeat the transmission limb sections of Figure \ref{fig:trans} but show the composition of the cloud particles through their individual contributions to the VMR of the grain bulk.
Perhaps the most surprising result is the mixing ratio of Fe[s] is generally higher than \ce{Mg2SiO4}[s] at the millibar pressure regions, with \ce{Mg2SiO4}[s] dominating at pressures higher and lower than $\approx$10$^{-2}$ bar. 
This is most probably due to the Kelvin effect (Section \ref{sec:KE}), where due to the high surface tension value of Fe[s], a larger supersaturation ratio is required to condense Fe[s] on the grain surfaces.
This behaviour is also seen in other hot Jupiter cloud microphysical studies that include the Kelvin effect \citep[e.g.][]{Powell2018, Powell2019}.
\ce{TiO2}[s] and \ce{Al2O3}[s] do not contribute significantly to the cloud particle bulk, only appearing at around 1 bar in small pockets of the east and west equatorial regions.

\section{Post-processing}
\label{sec:pp}

To post-process our HAT-P-1b GCM results we use the 3D Monte Carlo RT model gCMCRT \citep{Lee2022}.
We focus on comparing to the available transmission spectral data from \citet{Wakeford2013,Nikolov2014} and \citet{Sing2016} and Spitzer emission data from \citet{Todorov2010}.
We then focus on producing synthetic spectra for the NIRISS SOSS instrument mode onboard JWST owing to the ongoing observational program of \citet{Lafreniere2017}.

\subsection{Transmission spectra}
\label{sec:trans_spec}

\begin{figure*}
    \centering
    \includegraphics[width=0.49\textwidth]{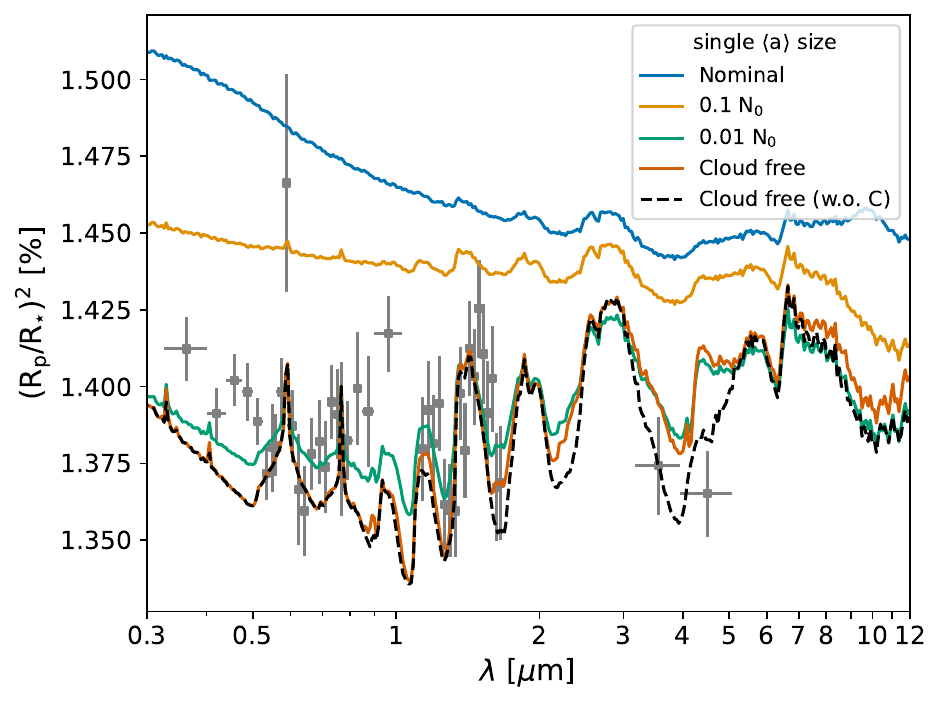}
    \includegraphics[width=0.49\textwidth]{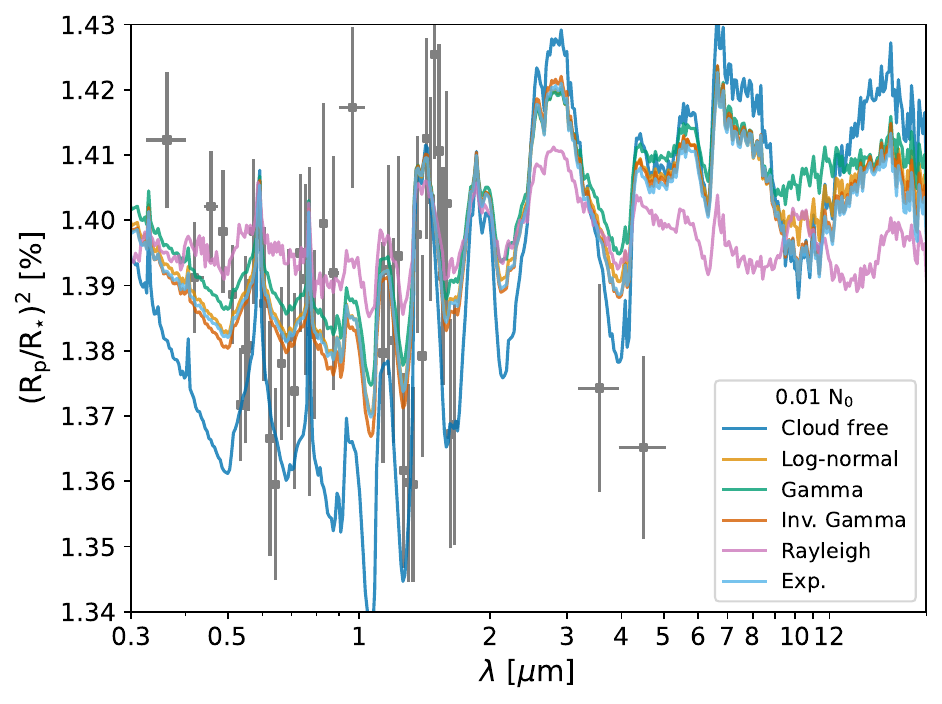}
    \caption{Transmission spectra produced from the GCM results. 
    Left: Assuming a single particle size at the mean value with different multiple factor of the total number density N$_{0}$.
    Right: Assuming a N$_{0}$ factor of 0.01 with different assumptions of the particle size distribution shape.
    Observational data (grey points with error bars) are taken from \citet{Wakeford2013,Nikolov2014} and \citet{Sing2016}.}
    \label{fig:trans_spec}
\end{figure*}

In Figure \ref{fig:trans_spec} we present the transmission spectra resulting from the GCM model.
From this, it is clear that the nominally produced cloud structures presented in the previous sections drastically over estimate the cloud opacity, resulting in a too flat spectra compared to the available HST and Spitzer data.  
We arbitrarily reduce the cloud number density by a factor of 0.01 to return the model to a more appropriate level. 
This is similar to the procedure used in \citet{Lines2018b}.
From the left hand side plot, we show that the cloud model can reduce the \ce{H2O} feature amplitudes and helps connect the HST WFC3 and STIS data in more consistent manner.
In the right hand side plot, we examine the effect of assuming different particle size distributions, derived from the moment solutions in the GCM (App. \ref{app:dist}). 
This shows the effect of each assumption on the spectrum. 
As expected, distributions that contain larger particle sizes, such as the Rayleigh distribution, flatten the spectrum more compared to distributions that prefer smaller particle sizes, such as the inverse gamma or exponential distribution.

Our post-processed GCM transit spectra are unable to fit the Spitzer points in both the cloud free and cloudy cases. 
Retrieval modelling performed in \citet{Barstow2017} suggest that a \ce{H2O} only atmosphere without carbon bearing species and a grey cloud deck can best explain the Spitzer values.
For one case in Figure \ref{fig:trans_spec}, we attempt to emulate this by removing carbon species from the spectra, which results in a smaller transit depth more in line with the observed Spitzer values.
Therefore, our nominal assumption of Solar Metallicity with carbon species in chemical equilibrium is possibly not the most accurate representation of the atmosphere, but this assumption enables greater ease and simplification of simulation within the coupled GCM and mini-cloud model.
Future JWST measurements in this wavelength region, most probably with G395H, will be highly enlightening for measuring the carbon content in the atmosphere.

\subsection{Emission spectra}

\begin{figure*}
    \centering
    \includegraphics[width=0.49\textwidth]{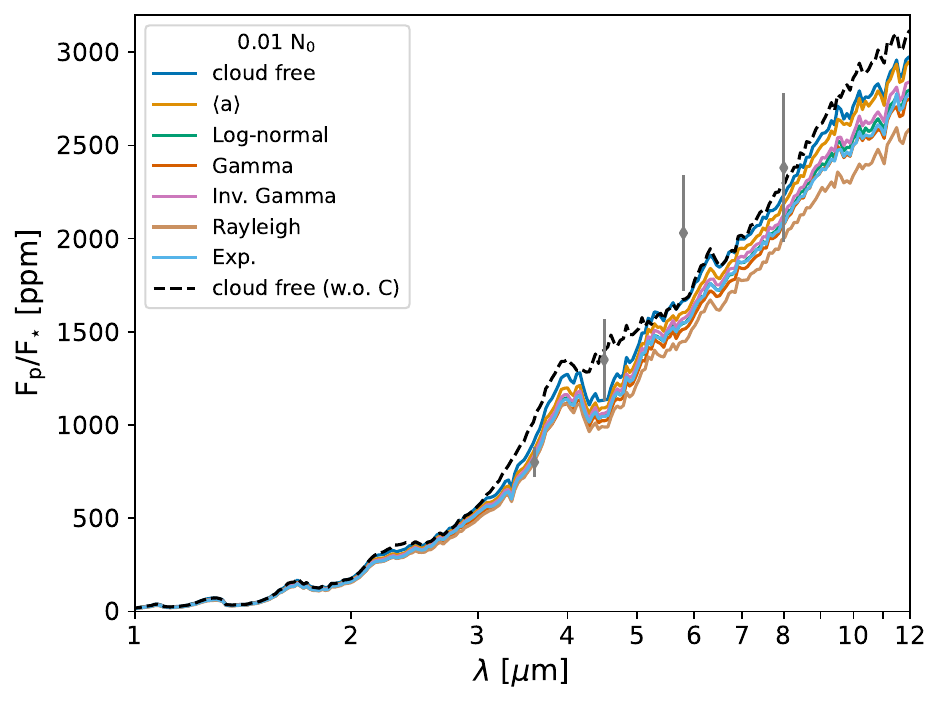}
    \includegraphics[width=0.49\textwidth]{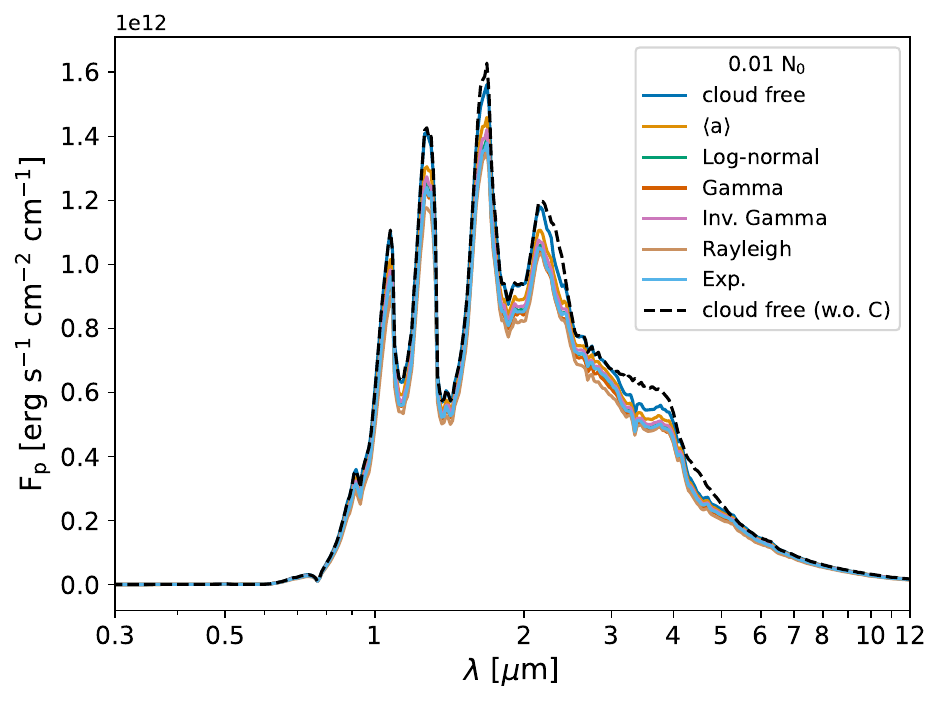}
    \caption{Dayside secondary eclipse spectra (Left) and planetary spectral flux (Right) produced using the cloudy GCM results. We assume a N$_{0}$ reduction factor of 0.01 which fit the transmission spectra better. Spitzer observational data is taken from \citet{Todorov2010}.}
    \label{fig:em_spec}
\end{figure*}

In Figure \ref{fig:em_spec} we present the secondary eclipse emission spectrum and dayside planetary flux from the GCM results including both the cloud free and cloudy spectra.
Clouds have a significant effect on the spectra at the redder end of the wavelength range, with each assumed distribution altering the shape of the spectra in different ways. 
For example, the Rayleigh distribution provides the largest cloud opacity, reducing the outgoing flux by several factors compared to other distributions.
From the spectral flux plot it is clear the cloud opacity reduces the outgoing flux across the entire wavelength regime.
All models are generally consistent with the 3.6 $\mu$m and 8 $\mu$m
photometric points, but underpredict the flux at 4.5 $\mu$m and 5.8 $\mu$m points. 
The model without carbon bearing species (black dashed point), suggested by the transmission spectra retrieval modelling (See discussion in Sect. \ref{sec:dis}), well fits the 4.5$\mu$m point.

\subsection{JWST NIRISS SOSS predictions}

\begin{figure*}
    \centering
    \includegraphics[width=0.49\textwidth]{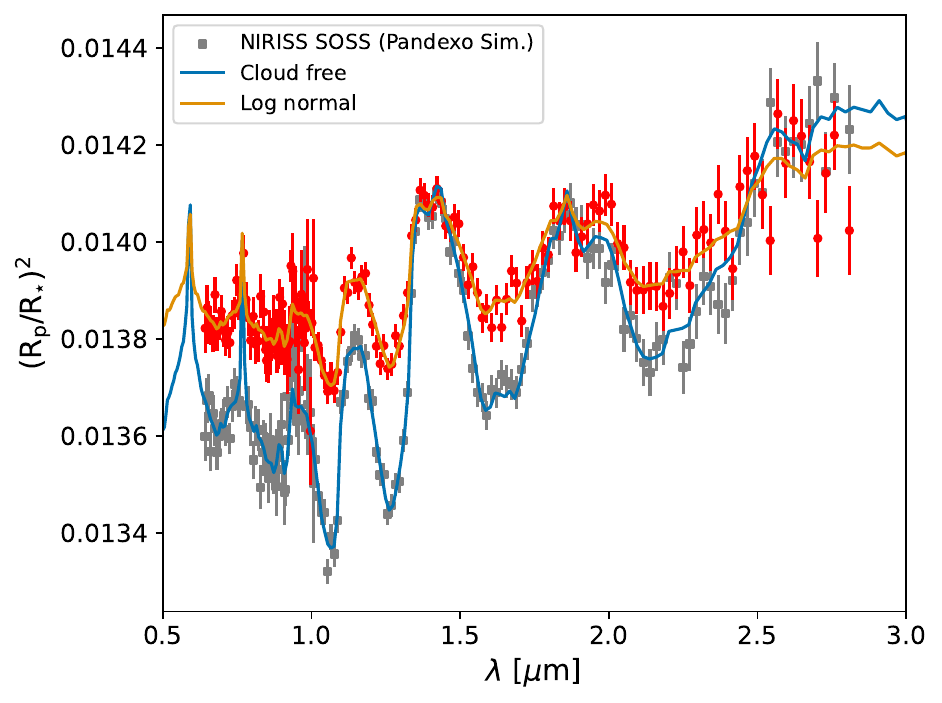}
    \includegraphics[width=0.49\textwidth]{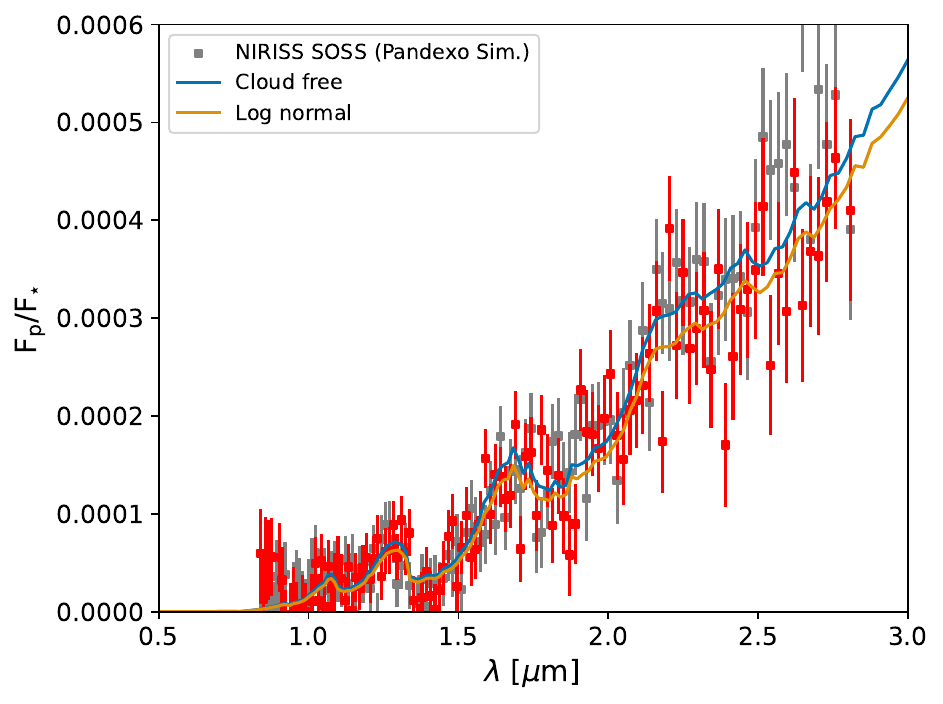}
    \caption{Simulated NIRISS SOSS transit spectra (left) and emission spectra (right) using Pandexo \citep{Batalha2017b}. 
    The cloud free (blue model and grey simulated points) and log normal (orange model and red simulated point) cases are compared.
    Cloudy affects on the spectrum are more discernible in transit rather than emission for this JWST mode. We bin the simulated points to R = 50 for clarity.}
    \label{fig:SOSS_spec}
\end{figure*}

The JWST GTO program 1201 \citep{Lafreniere2017} is scheduled to observe a single transit and eclipse of HAT-P-1b using NIRISS SOSS.
To put our GCM results in a practical context we produce simulated NIRISS SOSS transit and emission spectra using PandExo \citep{Batalha2017b}, presented in Figure \ref{fig:SOSS_spec}. 
We use the log-normal distribution case as representative of the cloudy spectra produced in Section \ref{sec:trans_spec}.
We use the same observational run parameters as in the \citet{Lafreniere2017} program as best as possible.

Our simulated spectra show that in transmission, the cloudy and non-cloudy case should be easily discernible, improving the precision from the HST data.
The shape of the spectrum at the potassium wings and flattening of the \ce{H2O} windows will provide ample evidence of the cloud coverage on HAT-P-1b as well as confirming the potential non-presence of potassium on this planet suggested by the HST STIS data \citep{Nikolov2014}.
However, according to Figure \ref{fig:trans_spec}, we suggest it may be difficult to differentiate between different size distributions even with this improved precision, but larger sized peaked distributions such as the Rayleigh distribution may be easily ruled out.

For the emission spectrum, our simulations show it will be harder to discern between the cloudy and non-cloudy case.
Evidence for dayside clouds may be sought at the reddest end of the spectrum, where the cloud model diverges from the cloud free case more.
Possibly, a combination of the SOSS data with the current Spitzer data would provide a much tighter constraint on the possible dayside cloud structures.

Overall, we suggest the prospects for in-depth characterisation of HAT-P-1b's atmosphere and accompanied cloud structures are extremely good. 
These upcoming data will be an excellent future testbed for modelling cloud formation in this hot Jupiter regime.

\section{Discussion}
\label{sec:dis}

In this study, we have used the moment method to evolve the integrated properties of the cloud particle size distribution across the simulation.
This makes an assumption that the cloud particle distribution will smooth across radius space and settle at an averaged rate. 
A more realistic depiction of the size distribution would be achieved using a `bin method' where each individual size bin of the distribution is integrated in time and has it's own settling rates.
The CARMA model \citep[e.g.][]{Gao2020} is an example of a microphysical bin model successfully used for hot Jupiter exoplanet cloud modelling \citep{Powell2018,Powell2019}.
If computationally feasible, coupling such a model to the GCM would be greatly enlightening as well as provide an insight into the most realistic distributions to assume for the moment method.

As stated earlier, we have assumed that the condensation of clouds does not alter the background atmospheric gas composition from Solar values, the oxygen abundance is not depleted self-consistently inside mini-cloud which has a large impact on the chemical composition of the atmosphere by reducing \ce{H2O} abundance and altering C/O ratio significantly \citep[e.g.][]{Helling2021,Helling2023}.
This may have a further feedback effect on the thermal structure of the atmosphere through changes in the gas phase composition.
We plan to take this into account in future studies by including the tracking of oxygen condensation inside mini-cloud, along with coupling a chemical equilibrium model to consistently calculate gas phase abundances for input to the RT scheme.

Related, we have assumed a solar metallicity atmosphere in our current simulations. 
An increase in metallicity will affect both the dynamical properties of the atmosphere, through an increase in the day/night temperature contrast and reduction in energy transport efficiency to the nightside of the atmosphere.
Cloud formation wise, increasing the metallicity will increase the thermal stability zone of cloud species to higher temperatures as well as increasing the available condensable mass.
This has the effect of compacting the clouds to thicker cloud layers deeper in the atmosphere.
In addition, a larger nucleation rate due to increased metallicity may allow a much larger population of smaller particles to reside in the atmosphere compared to the micron sized particles produced in this study.
The effect of metallicity on the 3D cloud structures is left to future work.

\subsection{Observational predictability}

As noted in the review in \citet{Zhang2020}, consistent microphysical cloud models coupled to GCMs tend to produce too flat transmission spectra with reductions of the cloud opacity by several 10s of factors required to reproduce spectral features \citep[e.g.][]{Lines2018}.
This problem persists in this study, suggesting that additional physical processes not considered are required to be added to mini-cloud before stronger predictions can be made using the model.

A key process that would reduce cloud number density and opacity would be coagulation, where cloud particles collide together in the atmosphere forming particles of larger sizes.
This processes is included in the bin models of CARMA \citep{Gao2020} but not in the moment method of the current study.
Possibly including the coagulation and fragmentation mechanism from \citet{Samra2022}, who also used a moment based method, would be a way of including coagulation into the current model.

In \citet{Barstow2017} retrieval modelling was performed on the available HST and Spitzer transmission spectra data for HAT-P-1b. 
They suggest a carbon free atmosphere is preferred, with only \ce{H2O} as the main gas phase absorber driven by the position of 4.5$\mu$m Spitzer photometric point. 
In Figures \ref{fig:trans_spec} and \ref{fig:em_spec} we also add this scenario in transmission and emission by arbitrarily removing the opacity of carbon bearing species from the post-processing routine. 
However, this wavelength range lies beyond the NIRISS SOSS limits meaning that, if possible, other instruments such as NIRSpec G395H would have to be used to investigate this scenario in detail.
Encouragingly, Spitzer transmission data has been shown to line up well with NIRSpec G395H data for the hot Saturn WASP-39b \citep{Ahrer2023}.

\subsection{Runtime}

In this study we report around 5 minutes per simulated day with the cloud-free model and with the correlated-k RT scheme \footnote{Using a server with Intel Xeon Gold 6130 CPU 2.10GHz CPUs}.
Coupled to mini-cloud using 48 processors with radiative-feedback we report an approximate maximum 15 minutes per simulated day, suggesting the addition of the cloud microphysics module adds approximately 3x the computational burden to the basic GCM model.
Compared to the previous models that used DIHRT \citep{Lee2016,Lines2018} which reported $\sim$hours of simulation per simulated day, mini-cloud offers a magnitude increase in computational efficiency and puts 3D microphysical cloud modelling within reach of most contemporary GCM modelling groups.
We suggest that runtime can be further improved by using more modern HPC infrastructure, which would put longer timescale evolution into reach.

\subsection{Convergence}

In \citet{Woitke2020} coupled the DRIFT formalisms in a time dependent manner to a 1D settling and diffusive scheme, DIFFUDRIFT, to model brown dwarf and exoplanet atmospheres.
They found that long timescales of $\approx$ years were required for convergence to occur.
Due to the similarity in methodology and set-up to from \citet{Woitke2020} to this study, we find it likely similar timescales are required from the GCM simulations with clouds for convergence (or rather statistical convergence in the GCM case). 
However, our current study also concurs with some of the conclusions found in \citet{Woitke2020}, namely,
\begin{itemize}
    \item Seed particle nucleation rates are very low due to the slower mixing from the deep regions to the \ce{TiO2} nucleating zones.
    \item As a result, larger particles form more easily and are confined to deep layers while only small particles can be lofted upwards beyond the main condensation fronts.
\end{itemize}
These conclusions are similar to the 1D CARMA microphysical bin model results at similar T-p conditions found in \citet{Powell2018,Powell2019}.

It is therefore likely the cloud structures presented in this study are not fully converged. 
However, we note the dynamics of GCM simulations in this regime in general is unlikely to be fully converged due to the extremely long timescales ($\sim$ 100,000 days) required for momentum exchange between the deep and upper atmosphere \citep[e.g.][]{Wang2020}.
Both of these problems combined suggest that a truly statistically converged model that couples microphysical clouds to GCMs is a highly challenging prospect, beyond the scope of current model capabilities.

\section{Summary and Conclusions}
\label{sec:con}

In this study, we present the open source microphysical cloud formation module, `mini-cloud', designed for time dependent modelling of cloud particles in exoplanet atmospheres.
Mini-cloud offers an intermediate methodology between full microphysical models and phase equilibrium cloud models.
In particular, mini-cloud is most suited for examining cloud microphysics for 3D GCM modelling of cloud structures in hot gas giant exoplanets where computational expediency is a key requirement.

We simulated the cloud structures of the hot Jupiter exoplanet HAT-P-1b, finding that extensive cloud coverage is expected. 
However, we significantly overpredict the number density of cloud particles by approximately two orders of magnitude, which is required to fit the present transit and dayside emission observational data. 
This suggests that additional physical processes that can reduce cloud number density not included in the model, such as coagulation, may be important considerations for future modelling efforts.

Our results overall suggest that consideration of the 3D dynamical transport of particles and gas in addition to temperature is important in setting the global cloud properties of the atmosphere. 
The interplay between mixing of condensable material, avenues of replenishment to the upper atmosphere play a key role in determining the overall cloud coverage properties, with polar and high latitude regions acting as a source of cloud particle nucleation and formation in the HAT-P-1b case.

Mini-cloud enables simulation of cloud microphysics for 1000s of simulated days rather than the previous studies 100 days or less.
Mini-cloud offers a useful intermediate method between the microphysical cloud models and phase equilibrium models currently in used for 3D GCM modelling of exoplanet atmospheres, and puts the study of cloud formation microphysics into reach of most contemporary exoplanet GCM modeller resources.

\section*{Acknowledgements}
E.K.H. Lee is supported by the SNSF Ambizione Fellowship grant (\#193448).
The HPC support staff at AOPP, University of Oxford is highly acknowledged.

\section*{Data Availability}
The mini-cloud and gCMCRT source codes are available on the lead author's GitHub: \url{https://github.com/ELeeAstro}.
All other data is available upon request to the lead author.

\clearpage



\bibliographystyle{mnras}
\bibliography{bib} 



\clearpage

\appendix

\section{Reconstructing size distributions from moment solutions}
\label{app:dist}

\begin{figure}
    \centering
    \includegraphics[width=0.49\textwidth]{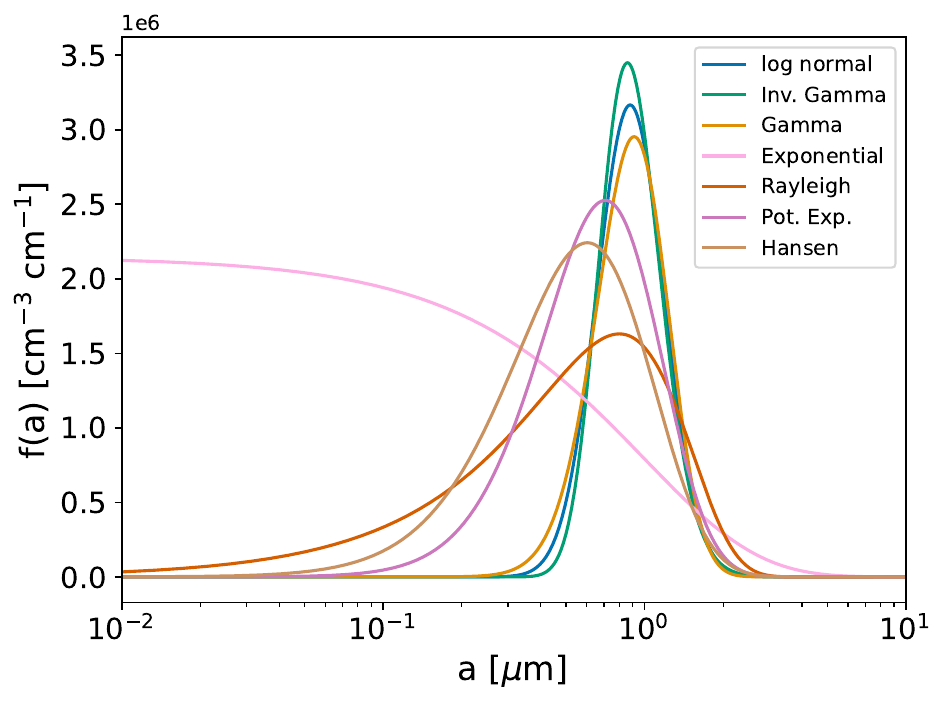}
    \caption{Reconstructed assumed cloud particle size distributions with the same moment solutions.}
    \label{fig:dist}
\end{figure}

Given specific shape information a size distribution can be reconstructed from the moment solutions by assuming a suitable arithmetic distribution shape to fit to the moment data. 
The relations between moments represent a mathematical property of the size-distribution, for example, the expectation value, E[a], is approximately equal to the mean cloud particle size $\left<a\right>$ = K$_{1}$/K$_{0}$, and the variance, Var[a], approximately equal to K$_{2}$/K$_{0}$ - (K$_{1}$/K$_{0}$)$^{2}$ or $\left<A\right>$/4$\pi$ - $\left<a\right>^{2}$.
In the following sections we provide a simple reference on various distributions that are commonly used across the literature.
In Figure \ref{fig:dist} we show the resultant reconstructed size-distributions for a mini-cloud test that has N$_{0}$ = 215 cm$^{-3}$, E[a] = 10$^{-4}$ cm and Var[a] = 9.14 $\cdot$ 10$^{-10}$ cm$^{2}$.

\subsection{Log-normal distribution}
The log-normal distribution is a ubiquitous assumed distribution for describing clouds in gas giant atmospheres \citep[e.g.][]{Ackerman2001}.
It is characterised by having the mean, mode and median the same value, with a broad distribution spread equally in log space across a given variance value.
The log-normal distribution is given by
\begin{equation}
    f(a) = \frac{N_{0}}{a\sigma\sqrt{2\pi}}\exp\left[-\frac{(\ln{a} - \mu)^{2}}{2\sigma^{2}}\right],
\end{equation}
where N$_{0}$ = K$_{0}$ [cm$^{-3}$] is the total cloud particle number density, $\mu$ the natural logarithm of the distribution mean and $\sigma$ the natural logarithm of the distribution variance.
$\mu$ is estimated through
\begin{equation}
    \mu = \ln{\left(\frac{E[a]}{\sqrt{\frac{Var[a]}{E[a]^{2}} + 1}}\right)}
\end{equation}
\begin{equation}
    \sigma = \sqrt{\ln\left(\frac{Var[a]}{E[a]^{2}} + 1\right)}
\end{equation}

\subsection{Inverse Gamma distribution}
The inverse gamma distribution is a common alternative to the log-normal distribution, characterised by an given by
\begin{equation}
    f(a) = \frac{N_{0}}{\Gamma(\alpha)}\beta^{\alpha}a^{-\alpha-1}\exp\left(-\frac{\beta}{a}\right),
\end{equation}
where the $\alpha$ and $\beta$ parameters are estimated from the two relations
\begin{equation}
    \alpha = \frac{E[a]^{2}}{Var[a]} + 2.0
\end{equation}
and
\begin{equation}
    \beta =  E[a](\alpha - 1)
\end{equation}
\subsection{Gamma distribution}
The gamma distribution is given by
\begin{equation}
    f(a) = \frac{N_{0}}{\Gamma(\alpha)}\beta^{\alpha}a^{\alpha-1}\exp\left(-\beta a\right),
\end{equation}
where the $\alpha$ and $\beta$ parameters are estimated from the two relations
\begin{equation}
    \alpha = \frac{E[a]^{2}}{Var[a]}
\end{equation}
and
\begin{equation}
    \beta = \frac{E[a]}{Var[a]}
\end{equation}

\subsection{Exponential distribution}
The exponential distribution is given by
\begin{equation}
    f(a) = N_{0}\lambda\exp(-\lambda a),
\end{equation}
where lambda is 
\begin{equation}
    \lambda = \frac{1}{E[a]}.
\end{equation}

\subsection{Rayleigh distribution}
The Rayleigh distribution is special form of the gamma distribution given by
\begin{equation}
    f(a) = \frac{N_{0}a}{\sigma^{2}}\exp\left(-\frac{a^{2}}{2\sigma^{2}}\right),
\end{equation}
with $\sigma$ given by
\begin{equation}
    \sigma = \frac{E[a]}{\sqrt{\pi/2}}.
\end{equation}
Although the Rayleigh distribution is simple to use, as a single parameter distribution it can struggle to represent well peaked distributions.

\subsection{Potential exponential distribution}
\citet{Helling2008} propose fitting moment solutions to a potential exponential distribution as a variant of the Gamma distribution
\begin{equation}
    f(a) = a^{B}\exp\left(A - Ca\right),
\end{equation}
with 
\begin{equation}
    B = \frac{2K_{1}K_{3} - 3K_{2}^{2}}{K_{2}^{2} - K_{1}K_{3}},
\end{equation}
\begin{equation}
    C = (2 + B)\frac{K_{1}}{K_{2}},
\end{equation}
and
\begin{equation}
    A = \ln K_{1} + (2 + B)\ln C - \ln\Gamma(2 + B).
\end{equation}

\subsection{Hansen distribution}
The Hansen distribution was proposed in \citet{Hansen1971} to fit observed cloud particle size distributions on Earth but has been used in Brown Dwarf and exoplanet atmospheric contexts \citep[e.g.][]{Hiranaka2016,Burningham2017}.
It follows an alternative formulation of the Gamma distribution
\begin{equation}
    f(a) = \frac{N_{0}}{\Gamma[(1 - 2\beta)/\beta]}(\alpha\beta)^{(2\beta - 1)/\beta}\alpha^{(1 - 3\beta)/\beta}\exp\left(-\frac{a}{\alpha\beta}\right),
\end{equation}
where $\alpha$ is the effective particle radius (Eq. \ref{eq:aeff}) and $\beta$ the effective variance of the particle size distribution, these can be estimated as 
\begin{equation}
    \alpha = \frac{K_{3}}{K_{2}},
\end{equation}
\begin{equation}
    \beta = \frac{K_{2}K_{4}}{K_{3}^{2}} - 1.
\end{equation}
Since the K$_{4}$ moment is not directly calculated in mini-cloud, it may be reasonable to approximate $\beta$ by using ratios of lower integer moments
\begin{equation}
    \beta = \frac{K_{1}K_{3}}{K_{2}^{2}} - 1.
\end{equation}

\bsp	
\label{lastpage}
\end{document}